\newcommand{\onecol}{true}
\newcommand{\comments}{false}

\ifx\onecol
	\documentclass[onecolumn,iop,numberedappendix]{emulateapj}
\else
	\documentclass[iop,numberedappendix]{emulateapj}
\fi

\usepackage{mathtools}
\usepackage{amsmath}
\usepackage{xcolor}
\usepackage{xparse}
\usepackage{bm,upgreek}

\newcommand{\HI}{H\textsc{i}}
\newcommand{\kpar}{\ensuremath{k_{||}}}
\newcommand{\kunit}{\ensuremath{h{\rm Mpc}^{-1}}}
\newcommand{\sbrk}{\ensuremath{S_\mathrm{break}}}
\newcommand{\N}{\ensuremath{\tilde{\mathcal{N}}}}

\ExplSyntaxOn
\NewDocumentCommand\vect{m}
{
	\commexo_vector:n { #1 }
}

\cs_new_protected:Npn \commexo_vector:n #1
{
	\tl_map_inline:nn { #1 }
	{
		\commexo_vector_inner:n { ##1 }
	}
}

\cs_new_protected:Npn \commexo_vector_inner:n #1
{
	\tl_if_in:VnTF \g_commexo_latin_tl { #1 }
	{% we check whether the argument is a Latin letter
		\mathbf { #1 } % a Latin letter
	}
	{% if not a Latin letter, we check if it's an uppercase Greek letter
		\tl_if_in:VnTF \g_commexo_ucgreek_tl { #1 }
		{
			\bm { #1 } % a Greek uppercase letter
		}
		{% if not, we check if it's a lowercase Greek letter
			\tl_if_in:VnTF \g_commexo_lcgreek_tl { #1 }
			{
				\commexo_makeboldupright:n { #1 }
			}
			{% none of the above, just issue #1
				#1 % fall back
			}
		}
	}
}

\cs_new_protected:Npn \commexo_makeboldupright:n #1
{
	\bm { \use:c { up \cs_to_str:N #1 } }
}

\tl_new:N \g_commexo_latin_tl
\tl_new:N \g_commexo_ucgreek_tl
\tl_new:N \g_commexo_lcgreek_tl
\tl_gset:Nn \g_commexo_latin_tl
{
	ABCDEFGHIJKLMNOPQRSTUVWXYZ
	abcdefghijklmnopqrstuvwxyz
}
\tl_gset:Nn \g_commexo_ucgreek_tl
{
	\Gamma\Delta\Theta\Lambda\Pi\Sigma\Upsilon\Phi\Chi\Psi\Omega
}
\tl_gset:Nn \g_commexo_lcgreek_tl
{
	\alpha\beta\gamma\delta\epsilon\zeta\eta\theta\iota\kappa
	\lambda\mu\nu\xi\pi\rho\sigma\tau\upsilon\phi\chi\psi\omega
	\varepsilon\vartheta\varpi\varphi\varsigma\varrho
}

\ExplSyntaxOff
\newcommand{\cpois}{\ensuremath{\vect{C}_{\rm FG}^{\rm Poiss}}}
\newcommand{\cclust}{\ensuremath{\vect{C}_{\rm FG}^{\rm Clust}}}

\shorttitle{EoR Point-Source Foregrounds}
\shortauthors{Murray et al.}

\begin{document}
	
	
	\title{An Improved Statistical Point-Source Foreground Model for the Epoch of Reionization}
	
	
	\author{S.~G.~Murray\altaffilmark{1,2}, C.~M.~Trott\altaffilmark{1,2} and C.~H.~Jordan\altaffilmark{1,2}}
	
%	\author{C. D. Biemesderfer\altaffilmark{4,5}}
%	\affil{National Optical Astronomy Observatories, Tucson, AZ 85719}
%	\email{aastex-help@aas.org}
%	
%	\and
%	
%	\author{R. J. Hanisch\altaffilmark{5}}
%	\affil{Space Telescope Science Institute, Baltimore, MD 21218}
	
	
	\altaffiltext{1}{ARC Centre of Excellence for All-sky Astrophysics (CAASTRO), }
	\altaffiltext{2}{International Centre for Radio Astronomy Research (ICRAR), Curtin University,  Bentley, WA 6102, Australia}
		

%%%%%%%%%%%%%%%%%%%%%%%%%%%%%%%%%%%%%%%

\begin{abstract}
We present a sophisticated statistical point-source foreground model for low-frequency radio Epoch of Reionization (EoR) experiments using the 21~cm neutral hydrogen emission line.
Motivated by our understanding of the low-frequency radio sky, we enhance the realism of two model components compared with existing models: the source count distributions as a function of flux density and spatial position (source clustering), extending current formalisms for the foreground covariance of 2D power spectral modes in 21~cm EoR experiments.
The former we generalise to an arbitrarily broken power-law, and the latter to an arbitrary isotropically-correlated field.
This paper presents expressions for the modified covariance under these extensions, and shows that for a more realistic source spatial distribution, extra covariance arises in the EoR window which was previously unaccounted for. Failure to include this contribution can yield bias in the final power spectrum and under-estimate uncertainties, potentially leading to a false detection of signal.
The extent of this effect is uncertain, owing to ignorance of physical model parameters, but we show that it is dependent on the relative abundance of faint sources, to the effect that our extension will become more important for future deep surveys. 
Finally, we show that under some parameter choices, ignoring source clustering can lead to false detections on large scales, due to both the induced bias and an artificial reduction in the estimated measurement uncertainty. 
\end{abstract}

\section{Introduction}
The EoR is a key era in the evolution of the Universe, in which the predominant neutral hydrogen component of the intergalactic medium is reionized due to the emerging activity of luminous sources. 
The fundamental importance of this epoch, as a bridge between large-scale cosmology and galaxy evolution, has solidified its statistical detection as a primary science goal of several current and upcoming instruments, such as the MWA \citep{Lonsdale2009,Tingay2013,Jacobs2016}; PAPER, \citep{Parsons2009}; LOFAR, \citep{vanHaarlem2013,Patil2016}; the LWA, \citep{Ellingson2009}; HERA, \citep{deBoer2016}; and the SKA \citep{Koopmans2015}.%{CHIME,PAPER,HERA,MWA,LOFAR,SKA}.

While optical surveys of the scant luminous sources of the EoR remain an interesting pursuit, a great deal of attention has been focussed on a statistical detection of the spatial signature of 21 cm emission arising from the patchy reionization process. Statistical detection, as opposed to the direct imaging that may be achievable with the SKA \citep{Koopmans2015}, offers the benefit of increasing detection signal-to-noise, but also provides a cosmological window on structure formation and evolution.
This emission pattern, redshifted into the low-frequency radio regime, offers a unique and powerful window into the heart of the interaction between large-scale-structure evolution and the emergence of luminous objects.
%That the signal itself is extremely weak necessitates a \textit{statistical} detection, rather than direct imaging (as may be possible with the Square Kilometre Array \cite{SKA}).
However, even statistical detection is beset with challenges due to systematic contamination of the signal with foreground emission (both Galactic and extragalactic), instrumental effects and other distortions (such as the effect of the ionosphere (Jordan 2017, submitted). 
It is commonly estimated that the foreground emission is 4-5 orders of magnitude brighter than the prized EoR signal \citep{Morales2009}.

It is no surprise then that much of the work being undertaken to detect the EoR concerns the unravelling of foreground emission from the true EoR signal. 
Commonly, this process takes one of three directions: the foregrounds can either be removed, avoided or suppressed. 
A useful diagnostic tool in this regard has been the 2D cylindrical power spectrum, and it also aids understanding these three general methods.

While the primary statistic employed for characterisation of the spatial EoR signature is the 1D isotropic power spectrum of \HI\ brightness temperature fluctuations, $P(k)$, the 2D power spectrum is a useful intermediate product. 
The cylindrical power spectrum (hereafter PS) bins modes in the plane of the sky, $k_\perp$, and modes parallel to the line of sight, $\kpar$.
This is important, because radio telescopes probe these modes in vastly different ways.
Perpendicular modes are dictated by baseline lengths and beam patterns, whereas parallel modes are dictated by frequency-based processes. 
This extends to the foreground emission, whose spectral structure (and thus parallel-mode structure) is entirely different from its angular distribution on the sky.
Thus, different systematic effects can be localised and dealt with on a 2D PS map. 

Of particular note is the fact that typical spectral structure for foreground sources is smooth and power-law-like. 
This confines the majority of foreground power to large parallel scales (low $\kpar$).
This feature opens up the way for the three aforementioned methods.
The foregrounds can be avoided by only averaging over high-$\kpar$ modes (the so-called EoR ``window"), in which the foreground power is low, but this may omit the incorporation of much useful information.
Furthermore it presumes we understand precisely where the foregrounds stop and the EoR window starts.
The foregrounds can be removed by attempting to fit each detected source with some spectral model (either with parametric models \citep[eg.][]{Bowman2009,Liu2009}, or with a blind component analysis \citep[e.g.,][]{Chapman2014}), but this runs the risk of residuals imitating the EoR signature.
The foregrounds could otherwise be suppressed, by assigning appropriate weights to modes in which the foregrounds are expected to dominate \citep[e.g.,][]{Liu2011}. 
Finally, some mixture of the approaches could be utilised.

In this paper we focus on developing statistical foreground models for use in a hybrid removal-suppression scheme, namely the \textit{Cosmological \HI{} Power Spectrum estimator} \citep[CHIPS;][hereafter T16]{Trott2016}.
While this scheme utilises foreground removal for the very brightest sources, for which accurate models can be fitted, it relies on a statistically consistent down-weighting mechanism to extract maximum information from the interferometric visibilities.
This has the advantage of statistically accounting for \textit{all} sources in the field of view, even those well below the detection threshold.
It achieves this through the inverse covariance matrix of the modes, which describes not only the uncertainty on a mode, but its correlation with others. 

Naturally, such a scheme requires a model for the distribution of foreground emission in order to derive its covariance. 
While the full model comprises Galactic emission and that of extragalactic sources, these are additive, and we focus 
here on the extragalactic point-source model.
Our reason for doing so is purely pragmatic -- realistic models of the Galactic emission (including spatially correlated structure) have already been proposed \citep[eg.][T16]{Jelic2010}, and the varying components are uncorrelated, providing opportunity to focus on each individually.
The model need only be statistical, and thus our aim in this paper is to present an improved statistical point-source foreground model in the context of the covariance of the point-source contribution to interferometric visibilities.

While the CHIPS scheme is quite sophisticated, several of its components are quite simplistic. 
In particular, it assumes both a single power-law source count distribution (as a function of flux density) and a uniform Poisson spatial distribution of sources on the sky.
The primary aims of this paper are to (i) propose generalisations of these simplistic models, (ii) derive the modified foreground covariance under these generalisations, and (iii) estimate the effect that ignoring these generalisations has on the averaged 1D power spectrum.

While we address in some measure all four of the relevant foreground model components -- the instrumental beam, the point-source SED, the source-count distribution and the spatial structure -- the latter-most component receives the most attention. 
This is due in part to the relative insensitivity of the results to other components, but also to the additional complexity a more complex spatial structure induces.

We note that previous work in the literature has broached the subject of spatially correlated foreground point-sources. 
In particular, the simulations of \cite{Jelic2010}, which include simple point-source clustering, are commonly used as test-beds for foreground mitigation schemes \citep[eg.][]{Chapman2014,Chapman2014a}, while more physically-motivated simulations are used in \cite{Sims2016}. 
However, the relatively small set of simulations presented are unable to yield the \textit{covariance} of foregrounds on all relevant scales.
In any case, an analytic model for the covariance is helpful as it is more efficient and aids the understanding of the features in the model.
An analytic treatment of point-source covariance in the presence of clustering was given in \cite{Liu2011}, however their results apply to the covariance in real-space.
We extend the analysis to the covariance of visibilities in Fourier-space, which is the natural space for interferometric observations, and in which we can include the chromaticity of the instrument and other realistic effects.

The paper is structured as follows. 
After some notational preliminaries in \S\ref{sec:prelim}, we will review the parts of the CHIPS scheme applicable to the point-source foreground covariance in \S\ref{sec:chips}. 
From there we turn to each of our target simplifications in turn -- the source-count distribution in \S\ref{sec:source_counts} and the spatial distribution in \S\ref{sec:clustering}. 
These sections develop the necessary mathematical machinery required to determine expected effects of our extensions with respect to synthetic data, which we present in \S\ref{sec:application}, before offering some concluding remarks in \S\ref{sec:conclusion}. 

We note that all plots in this paper were produced with \textsc{matplotlib} v2.0.0 \citep{Hunter2007}, with frequent use of the \textsc{numpy}, \textsc{scipy} \citep{Perez2007} and \textsc{ipython} \citep{vanderWalt2011} libraries.

\section{Notation and Preliminaries}
\label{sec:prelim}
\subsection{Notation}
Throughout, we use bold upright lower-case characters (Latin or Greek) to denote vectors (eg. $\vect{u}$ or $\vect{\theta}$), while bold upright upper-case characters denote matrices (notably the covariance matrix $\vect{C}$).
Statistical variables will be denoted by upper-case characters with an over-tilde (eg. $\tilde{\mathcal{N}}$).
Furthermore, sample means will be denoted by angled brackets, eg. $\langle \tilde{V} \rangle$, whereas population means will be denoted with an over-bar, eg. $\bar{n}$. 
Variance and covariance operators will be denoted by ${\rm Var}$ and ${\rm Cov}$ respectively.

\subsection{Fourier Transforms}
\label{sec:prelim:fourier}
Fourier-transforms (hereafter FTs) are defined with an unfortunate multiplicity of conventions in the literature, 
%and in this paper we incorporate results from two different fields -- cosmology and interferometry -- which utilise alternate conventions. 
and in this paper it is convenient to use a more flexible definition.
Following the parameterisation of the \textsc{mathematica} software\footnote{Found at \url{http://mathworld.wolfram.com/FourierTransform.html}}, the continuous $n$-dimensional FT forward/inverse pair can be respectively written 
\begin{align}
	F(\vect{k}) &= \sqrt{\frac{|b|}{(2\pi)^{1-a}}}^n \int f(\vect{r}) e^{-i b\mathbf{k}\cdot\mathbf{r}} d^n\vect{r}, \\
	f(\vect{r}) &= \sqrt{\frac{|b|}{(2\pi)^{1+a}}}^n \int F(\vect{k}) e^{+i b\mathbf{k}\cdot\mathbf{r}} d^n \vect{k},
\end{align}
where we shall denote the leading factors in each by  $\Psi^+_{a,b}$ and $\Psi^-_{a,b}$ respectively. 
Here $a$ and $b$ can be chosen arbitrarily, and fully specify the Fourier convention. Typically, interferometry uses the convention $(a,b) = (0,2\pi)$, however we shall see that it can be useful to keep these values arbitrary.
%while cosmology uses the convention $(a,b)=(1,1)$  \citep[eg.][though this is modified by a factor of the box-size for finite samples]{Peacock1999}. 
 Throughout, we use the operator $\mathcal{F}_{a,b}$ to denote a general continuous FT with specified convention, and reserve hat notation, eg. $\widehat{W}$, explicitly for a FT with the $(0,2\pi)$ convention.

In numerical calculations, only a discrete sampling of points is recorded, and the continuous FT must be approximated by a discrete FT (DFT). 
Such an operation can be conveniently handled using linear algebra, in which the 1D FT of a vector $\vect{x}$ is written $\psi \vect{F}\vect{x}$.
Here, $\psi$ is a normalisation which we shall specify in a moment, and $\vect{F}$ is a unitary Vandermonde matrix:
\begin{equation}
F_{mn} = \frac{1}{\sqrt{N}} \exp(-2\pi i mn/N),\ \ \ \ m,n \in (0,1,...,N-1),
\end{equation}
with $N$ the length of the data vector $\vect{x}$.
The matrix $\vect{F}$ has the property $\vect{F}^\dagger \vect{F} = \vect{F}\vect{F}^\dagger  = \vect{I}$, where $\vect{F}^\dagger$ encodes the inverse Fourier operation.

We typically require the DFT to approximate a continuous FT, and the two can be related by requiring that
\begin{align}
	\mathcal{F}_{a,b} &\rightarrow \psi^+ = \frac{ L}{\sqrt{N}} \Psi^+_{a,b}, \\
    \mathcal{F}_{a,b}^{-1} &\rightarrow \psi^- = \frac{2\pi \sqrt{N}}{bL} \Psi^-_{a,b},
\end{align}
where $L$ is the physical length of the discrete sample. The physical modes, $k$, that are measured in such a transform are 
\begin{equation}
k = \frac{2\pi}{b} \frac{m}{L}, \ \  m \in (-N/2,..., N/2),
\end{equation}
which relates the size of the Fourier-space box to the real-space box by $L_k = 2\pi N/L$.

\subsection{Interferometry}
Unless otherwise stated, we reserve the 2-vector $\vect{u}$ for baseline displacement in units of wavelength, 
\begin{equation}
	\label{eq:baselines}
	\vect{u} = \vect{x}/\lambda.
\end{equation}
Likewise, we reserve the 2-vector $\vect{l}\in(-1,1)$ for the sky co-ordinate
\begin{equation}
	\vect{l} = \sin\vect{\theta},
\end{equation}
with $\vect{\theta}$ the angle between the co-ordinate and zenith.

With these, in the flat-sky approximation, which we will generally assume throughout for simplicity, an interferometric visibility is defined as the Fourier-transform of the beam-attenuated sky brightness,
\begin{equation}
	\label{eq:visibility}
	V(\vect{u},\nu) = \int S(\vect{l},\nu) B(\vect{l},\nu) e^{-2\pi i \vect{u}\cdot\vect{l}}d\vect{l} \equiv \mathcal{F}_{0,2\pi} \left(S_\nu B_\nu \right)
\end{equation}
where $S$ is here the total flux density in a given direction, $B$ is the frequency-dependent beam attenuation and the integral is over the entire valid space of $\vect{l}$, i.e., $|\vect{l}|\leq 1$.

\subsection{Cosmology}
The cylindrical PS is defined at cosmological modes perpendicular, $k_\perp$, and parallel, $\kpar$, to the line-of-sight. 
The units of $k$ throughout are $\kunit$, where $h$ is the Hubble parameter. 
We detail the conversion of radio-astronomy units (Jy, Hz, sr) to cosmological units (mK, Mpc, $h$) in Appendix \ref{app:cosmo}.

For all cosmological calculations in this paper we use the parameters from \cite{PlanckCollaboration2015}, i.e. a flat Universe with $h=0.677$ and $\Omega_m=0.307$.

\section{The CHIPS Framework}
\label{sec:chips}
The basic idea behind the CHIPS algorithm is threefold:
\begin{enumerate}
	\item Individually model and peel all sources above some brightness threshold, $S_{\rm max}$.
	\item Using statistical models of the remaining foregrounds, estimate the covariance between 3D modes of the measured power spectrum.
	\item Consistently down-weight modes using their inverse covariance \citep{Liu2011} in the process of final averaging to a spherically-symmetric 1D power spectrum.
\end{enumerate}
In this section, we briefly review the mathematical framework presented in T16 concerning the latter two of these steps, in order to contextualise the remainder of the modelling in this paper.

\subsection{Inverse-Covariance Formalism}
\label{sec:chips:inverse_covariance}
We first note that here we ignore the edge-gaps present in the band-pass of the MWA -- a problem dealt with explicitly in T16 using Least-Squares-Signal-Analysis (LSSA). 
Our purposes in this paper are tangential to such instrument-dependent subtleties, thus for the sake of simplicity we assume a perfectly-sampled spectral window. This allows us to replace the operator $\mathcal{H}$ found throughout \S4 of T16 with the corresponding raw Fourier kernel matrix $\vect{F}$ (cf. \S\ref{sec:prelim:fourier}).

The CHIPS model only accounts for covariance between frequency modes, neglecting covariance between on-sky modes, and we follow suit here. 
In this case, we assume that the distribution of flux densities on the sky for any given frequency forms a (potentially correlated) Gaussian distribution, and the visibilities (see Eq. \ref{eq:visibility}) as a function of frequency are drawn from a complex normal:
\begin{equation}
	\tilde{V}(\vect{u},\nu) \sim \mathcal{C}\mathcal{N} \left\{\bar{V}(\vect{u},\nu), \vect{C}_u(V_{\nu'},V_{\nu''})\right\},
\end{equation}
where $\vect{C}$ is the foreground covariance between visibilities on the same baseline at varying frequencies, which we will attempt to model in this paper, and henceforth denote as $\vect{C}_{\rm FG}$.

A final FT over $\nu$ must be performed to generate the visibility in Fourier space, from which the final power spectrum is formed. 
We assume that all unit conversions from the natural units to cosmological units (cf. Appendix \ref{app:cosmo}) will be handled in a final step, and thus we find that the normalisation required to approximate the continuous FT here is $\psi = d\nu$, i.e. the frequency bin width.

Using the identity that if $\vect{y} = \vect{F}\vect{x}$ then ${\rm Cov}(\vect{y}) = \vect{F}^\dagger {\rm Cov}(\vect{x}) \vect{F}$, we find that the covariance of Fourier-space visibilities is ${\rm Cov}(V_u(\eta)) = d\nu^2 \vect{F}^\dagger \vect{C}_{\rm FG} \vect{F}$. Since the FT of a Gaussian random variable is also Gaussian, the distribution of $V_u(\eta)$ is entirely known. For a Gaussian variable, the covariance of its square is twice the square of its covariance, so we may write
\ifx\onecol
  \begin{equation}
      \label{eq:power_covariance}
      {\rm Cov}(P_u(\eta)) \equiv \vect{C}_P(u,\eta) = 2 d\nu^4 \left[\vect{F}^\dagger \vect{C}_{\rm FG} \vect{F} \right]^2 = 2 d\nu^4 \vect{F}^\dagger \vect{C}_{\rm FG}\vect{C}_{\rm FG} \vect{F}.
  \end{equation}
\else
  \begin{align}
      \label{eq:power_covariance}
      {\rm Cov}(P_u(\eta)) \equiv \vect{C}_P(u,\eta) &= 2 d\nu^4 \left[\vect{F}^\dagger \vect{C}_{\rm FG} \vect{F} \right]^2 \nonumber \\ 
      &= 2 d\nu^4 \vect{F}^\dagger \vect{C}_{\rm FG}\vect{C}_{\rm FG} \vect{F}.
  \end{align}
\fi
We will use the diagonal of this quantity (i.e. the variance) to determine the expected signal-to-noise of a given mode.

Using a maximum-likelihood (ML) estimator, the estimate of the 1D $P_k$ is given by (cf. Eq. 44 of T16)
\begin{equation}
	P_k = \frac{\sum_{i\in k} \vect{C}_P(\eta)_{,i} \vect{p} }{{\rm tr}(D^\dagger \vect{C}_P(\eta)_{,i}D)},
\end{equation}
where $D$ represents the operation of binning 2D modes into 1D annuli, and the syntax, ${}_{,i}$, denotes a projection on the bin including parameter covariances.
Clearly, the normalisation factor $2d\nu^4$ is cancelled in the averaging.

In principle, the entire framework is dependent only on an accurate model for the covariance of visibilities, $\vect{C}_{\rm FG}$, and we now turn to deriving this quantity.

\subsection{The CHIPS Point-Source Foreground Model}
\label{sec:chips:review}
In the absence of redshift information, sources retain two key properties at a given frequency $\nu$ -- a flux density $S(\nu)$ and a 2D sky position $\mathbf{l}$. 
Thus, the statistical description of the sources is completely defined by three functions: (i) a source-count distribution, $dN/dS(\nu)$, (ii) a spectral-energy distribution (SED), $S(\nu)$ and (iii) a spatial distribution, $dn/d\mathbf{l}(\nu)$.

Current models for each of these functions are surprisingly simple. The source-count distribution is assumed to be a power-law:
\begin{equation}
	\label{eq:source_counts_pl}
	\frac{dN}{dS}(\nu) = \alpha_{\nu} \left(\frac{S}{\rm Jy}\right)^{-\beta}  \ \   {\rm Jy}^{-1}{\rm sr}^{-1},
\end{equation} 
where measurements at $\nu_0\sim150$~MHz \citep{Intema2011} give $\alpha_{\nu_0} \sim 4100\  {\rm Jy}^{-1}{\rm sr}^{-1}$ and $\beta \sim 1.59$. 
The SED of every source is assumed to be a power-law over the frequencies covered in a single observation:
\begin{equation}
	\label{eq:sed}
	\frac{S(\nu)}{S(\nu_0)} = \left(\frac{\nu}{\nu_0}\right)^{-\gamma},
\end{equation}
with a uniform value of $\gamma=0.8$ at $\nu_0\sim$150 MHz.
Finally, the spatial distribution is taken to be a Poisson process, i.e. a statistically uniform distribution across the sky.
We now turn to reviewing the derivation of the covariance of interferometric visibilities from these point-source foregrounds alone (as performed by T16).

\subsection{The CHIPS Point-Source Foreground Covariance}
\label{sec:review:cov}
Recall that the measured visibility is given by
\begin{equation}
	\tilde{V}(\vect{u},\nu) = \int \tilde{S}_T(\vect{l},\nu) B(\vect{l},\nu) e^{-2\pi i \vect{u}\cdot\vect{l}}d^2\vect{l} \ \ [{\rm Jy}],
\end{equation}
where we have now indicated that the visibility is a random variable, due to the inclusion of the random sky brightness $\tilde{S}_T$. 
According to the simple models introduced in the previous section, $\tilde{S}_T$ is given by
\begin{equation}
	\label{eq:total_flux_density}
	\tilde{S}_T(\nu)  = \int_0^{S_{\rm max}f_0^{-\gamma}} S \tilde{\frac{dN}{dS}}(\nu) dS \ \ [{\rm Jy}^{-1}{\rm sr}^{-1}],
\end{equation}
where $f_0 = \nu/\nu_0$, and $\nu_0$ is the reference frequency for the physical models.
A subtle point that entered into Eq. \ref{eq:total_flux_density} is that the integration limit involves a peeling flux $S_{\rm max}$ at a given frequency. The peeling flux is some flux density above which sources in the field are individually modelled and subtracted. In practice, the sources that are peeled are determined by their mean flux density over the bandwidth of the measurement. This implies that the number of sources in the field is identical at all flux densities (i.e. no source is peeled in one channel but not another). Assuming a \textit{universal} SED for all sources, this means that the peeling limit is simply modified according to the SED. 
Hereafter we shall denote $S_{\rm max}f_0^{-\gamma}$ simply as $S_{\rm max}^\nu$.

Another subtlety that requires explanation is to notice that an interferometric visibility measurement over a given frequency bandwidth $\Delta \nu$ will by necessity evaluate $\tilde{V}$ at a single vector $\vect{u}$, though in detail at each frequency, the physically measured $\vect{u}$ vary slightly due to Eq. \ref{eq:baselines}, where $\vect{x}$ are the constant physical baselines. Thus we arbitrarily choose a reference frequency within our bandwidth -- nominally $\nu_{\rm low}$, the lowest measured frequency -- and derive visibilities for each frequency in terms of the reference scales $\vect{u}$: $\vect{u}_\nu = \nu \vect{u}/\nu_{\rm low} \equiv f \vect{u}$. 

Including this information into the visibility yields
\begin{equation}
		\tilde{V}(\vect{u},\nu) = \int \int_0^{S_{\rm max}^\nu} S \tilde{\frac{dN}{dS}}(\nu) B(\vect{l},\nu) e^{-2\pi i f \vect{u}\cdot\vect{l}}dS d^2\vect{l},
\end{equation}

We now set out to evaluate the covariance of visibilities between frequencies, which we will denote $\cpois$. 
We approach this rather carefully, and in a more general manner than typically necessary so as to create a framework for our derivations in \S\ref{sec:clustering}.
To begin, we note the statistical identity
\begin{equation}
	\label{eq:general_cov_stat}
	{\rm Cov}\left(\sum X_i, \sum Y_i\right) = \sum_i \sum_j {\rm Cov}(X_i,Y_j),
\end{equation}
and that our double-integral is an example of such a sum of variables. 
In particular, the sum is over small voxels in the space of $(\vect{l},S)$, and thus the covariance is the sum of covariances of all pairs of voxels. 
We begin by imagining the voxel grid at $\nu_0$ and noting that each voxel contains a number of sources 
\begin{equation}
    \label{eq:cell_num}
    \tilde{N_i} \sim {\rm Poiss}\left(\frac{dN}{dS} \Delta S \Delta^2 \vect{l}\right).
\end{equation}
At a different frequency $\nu'$, each source is scaled equivalently by the universal SED ${f'_0}^{-\gamma}$. 
We imagine scaling the size of each voxel by the same factor so that the number of sources in each voxel remains constant, but the brightness of the voxel is scaled such that $S'_i = {f'_0}^{-\gamma} S_i$. 
The covariance between arbitrary voxels $i$ and $j$ is then
\begin{align}
 \vect{C}_{ij} = {\rm Cov}&\left[ {f'_0}^{-\gamma} S_i \tilde{N}_i B'_i e^{-2\pi i f' \vect{u}\cdot\vect{l}_i}, \right. \nonumber \\
  & \left. {f''_0}^{-\gamma} S_j \tilde{N}_j B''_j e^{-2\pi i f'' \vect{u}\cdot\vect{l}_j}\right] \nonumber.
\end{align}
We can extract the deterministic terms by noting that ${\rm Cov}(aX,bY) = ab^\dagger {\rm Cov}(X,Y)$, to retrieve
\begin{equation}
\vect{C}_{ij} = \left({f'_0 f''_0}\right)^{-\gamma} S_i S_j B'_i B''_j e^{-2\pi i \vect{u}\cdot(f'\vect{l}_i-f''\vect{l}_j)} {\rm Cov}\left[\tilde{N}_i,\tilde{N}_j\right]
\end{equation}
Since the counts are independently Poisson distributed, the final covariance factor simply reduces to $\bar{N} \delta_{ij}$, with $\delta_{ij}$ the Kronecker-delta, and the mean counts given in Eq. \ref{eq:cell_num}.
The double-sum over these covariance pairs reduces to a single sum due to the $\delta_{ij}$, and we abuse notation by re-using $i$ and $j$ for spatial and flux density co-ordinates to arrive at
\begin{equation}
    \cpois =  \left({f'_0 f''_0}\right)^{-\gamma}  \sum_i^{n_l}   B'_i B''_i e^{-2\pi i f_\nu \vect{l}_i \cdot \vect{u}} \Delta^2 \vect{l} \sum_j^{n_S} S_j^2 \frac{dN}{dS} \Delta S
\end{equation}
where $f_\nu = f'-f''$.
Finally, we allow the voxels to become infinitesimal, yielding
\ifx\onecol
\begin{equation}
	\label{eq:general_simple_covariance}
	\vect{C}_{\rm FG}^{\rm Poiss} =  (f'_0f''_0)^{-\gamma} \int B' B''e^{-2\pi i f_\nu \vect{u}\cdot\vect{l}} d^2\vect{l} \int_0^{S_{\rm max}} S^2 \frac{dN_0}{dS} dS. \equiv  (f'_0f''_0)^{-\gamma} \mu_2 \int B' B''e^{-2\pi i f_\nu \vect{u}\cdot\vect{l}} d^2\vect{l} ,
\end{equation}
\else
\begin{align}
\label{eq:general_simple_covariance}
\vect{C}_{\rm FG}^{\rm Poiss} &=  (f'_0f''_0)^{-\gamma} \int B' B''e^{-2\pi i f_\nu \vect{u}\cdot\vect{l}} d^2\vect{l} \int_0^{S_{\rm max}} S^2 \frac{dN_0}{dS} dS. \nonumber \\
&\equiv  (f'_0f''_0)^{-\gamma} \mu_2 \int B' B''e^{-2\pi i f_\nu \vect{u}\cdot\vect{l}} d^2\vect{l} \ \ \ [{\rm Jy}^2] ,
\end{align}
\fi
where $\mu_n$ is the $n^{th}$ moment of the source-count distribution:
\begin{equation}
	\label{eq:mu}
	\mu_n = \int S^n \frac{dN}{dS}dS \ \ \ [{\rm Jy}^n {\rm sr}^{-1}].
\end{equation}

It is worth mentioning some features that this model prediction contains. Firstly, the form of the source counts enters purely in a separate integral over flux density, and therefore their form affects the normalisation of the covariance, but not its structure (either in $\vect{u}$ or $\nu$). Since the inverse-covariance weighting method of foreground suppression is only sensitive to differences in weights across these scales, it is thus insensitive to the form of the source counts. We note that this strictly arises due to our assumption of a universal SED, which in reality is unjustified. We defer the incorporation of scatter in the SED to future work, but expect its influence to be minimal.

We also note that while there is a smooth frequency-dependence in the factor $(f'_0f''_0)^{-\gamma}$, which may give rise to power at very low $\kpar$, there is a more complex interaction between frequency and $\vect{u}$ in the exponent of the FT of the beam, and it is this interaction which gives rise to the so-called foreground `wedge'. Finally, the \textit{variance} of visibilities is simply the brightness-normalised integral over the beam squared:
\begin{equation}
 {\rm Var} \left[\tilde{V}\right] =  \mu_2 \int B^2 d^2\vect{l} \ \ \  [{\rm Jy}^2],
\end{equation}
which is clearly a constant.

Before proceeding, we note two approximations that can be (and are) made in this formalism that might otherwise escape notice. 
Firstly, we have already covertly approximated the spherical sky as a flat Euclidean space, by equating a uniform distribution of sources across the sky with a uniform distribution in $\vect{l}$. 
This is really just approximating $\vect{l}$ as equal to $\vect{\theta}$, which is a valid approximation close to the zenith.
To further utilise this approximation in order to simplify the covariance integral, one may assume that $\vect{l}$ is an infinite Euclidean space, i.e. it has no boundary at $|\vect{l}|=1$. 
While this is clearly unphysical, the presence of the attenuating beam in the integral ensures that for realistic telescopes, the contribution of high-$|\vect{l}|$ patches is negligible.
Thus the approximation is valid.
We follow the CHIPS methodology in adopting these approximations for ease of comparison, while noting that a more rigorous derivation would involve spherical harmonics \citep{Shaw2014,Liu2016}.
We leave such a derivation to future work.

Under these approximations, if we further assume that the beam is circularly symmetric, the integral becomes a Hankel transform,
\begin{equation}
	\label{eq:poisson_gaussian_covariance}
	\vect{C}_{\rm FG}^{\rm Poiss} = 2\pi (f'_0f''_0)^{-\gamma}  \mu_2 \int_0^\infty  l B'B'' J_0(2\pi u l)  \ dl,
\end{equation}
where $J_0$ is the zeroth-order Bessel function of the first kind.

Throughout this paper, we shall predominantly consider toy models that employ a circularly symmetric frequency-dependent Gaussian beam, $B(r) = \exp(-r^2/2\sigma_\nu^2)$, for which the final integral can be easily solved, to give
\begin{equation}
		\label{eq:poisson_gaussian}
			\vect{C}_{\rm FG}^{\rm Poiss} =  2\pi (f'_0f''_0)^{-\gamma}\mu_2 \Sigma_\nu^2 \exp(-2\pi^2 u^2 f_\nu^2 \Sigma_\nu^2),
\end{equation}
where 
\begin{equation}
	\Sigma_\nu^2 = \frac{\sigma''^2 \sigma'^2}{\sigma''^2 + \sigma'^2} \ \ \  [{\rm sr}].
\end{equation}
We note that while we use this simple form to illustrate features of our model in this paper, in practice the more general Eq. \ref{eq:general_simple_covariance} is used.

\section{A Generalised Source Count Model}
\label{sec:source_counts}
In the Poisson covariance model derived in T16 and examined in the previous section, the form of the source counts are important in setting the overall amplitude of the foreground covariance (and thus the expected foreground power). An incorrect model will mis-estimate the global level of power in foregrounds, and therefore provide inaccurate error-bars on the final 1D PS. 
Thus, using a high-fidelity source count model is somewhat important, however, in terms of the current framework, $\mu_2$ is dominantly affected by bright sources, unless the faint sources are much more abundant than we expect.
Since we can model bright sources accurately, it would appear that our current model is entirely adequate. 

%Nevertheless, perhaps a more important aspect of the inverse-covariance formalism is the assignment of different weights to different modes via the structure of the covariance. In this respect, the Poisson covariance model has no dependence on the source counts, which affect only the global normalisation. 
We find however that the introduction of clustering introduces a role for the source counts in modifying the structure of the covariance (cf. \S\ref{sec:clustering}).
In this case, an accurate source count model will be required to a much fainter limit and so developing a simple and well-defined model for the source counts becomes important.

The single power-law model for source counts is simple, but gives a remarkably good description of observations over observed flux density ranges at low frequencies \citep{Franzen2015,Williams2016}.
However, it is precisely this well-observed range that we expect to be peeled from the visibilities before an EoR analysis. Accordingly, it is the \textit{low} flux density regime we are interested in modelling. 
At the lowest flux densities, observations hint at a break in the power law, situated at $S \sim 6$ mJy \citep{Williams2016}. 
This break is thought to be the result of a transition between the regime in which AGN dominate the counts, to that of star-forming galaxies.
In addition, at some yet-fainter threshold another break must occur, in which the source counts turn over, to ensure a finite number of sources in the Universe (i.e. the low-$S$ slope must be $>-1$). 
It is likely that this turnover is quite sharp, and corresponds to the inability to form galaxies in low-mass halos. 

In this section, we briefly outline a more general model for the source counts -- namely an arbitrarily broken power-law -- flexible enough to approximately describe most possible future measurements.

\subsection{The Broken Power-Law Model}
Our choice to use a broken power-law is primarily driven by simplicity. 
While it offers discontinuous derivatives, and therefore is a poor choice in terms of a physical distribution, we expect that the \emph{statistical} properties of the true distribution can be arbitrarily well-recovered by a broken power-law, while at the same time presenting as an easily-integrable function.
It may be argued that a combination of various power-law populations is better modelled as a weighted sum of different power-laws, to which we in principle agree.
However the broken power-law model is better able to deal with anomalies in this structure, and furthermore has precedent in the literature \citep[eg.][]{Franzen2015}.
In any case, any viable parameterisation of the source counts is able to be swapped in to our formalism quite effortlessly, and the broken power-law provides a useful first look at some of the broad effects of the source counts.

\begin{figure}
\centering
\includegraphics[width=\linewidth, trim=0.7cm 0cm 0.7cm 0cm]{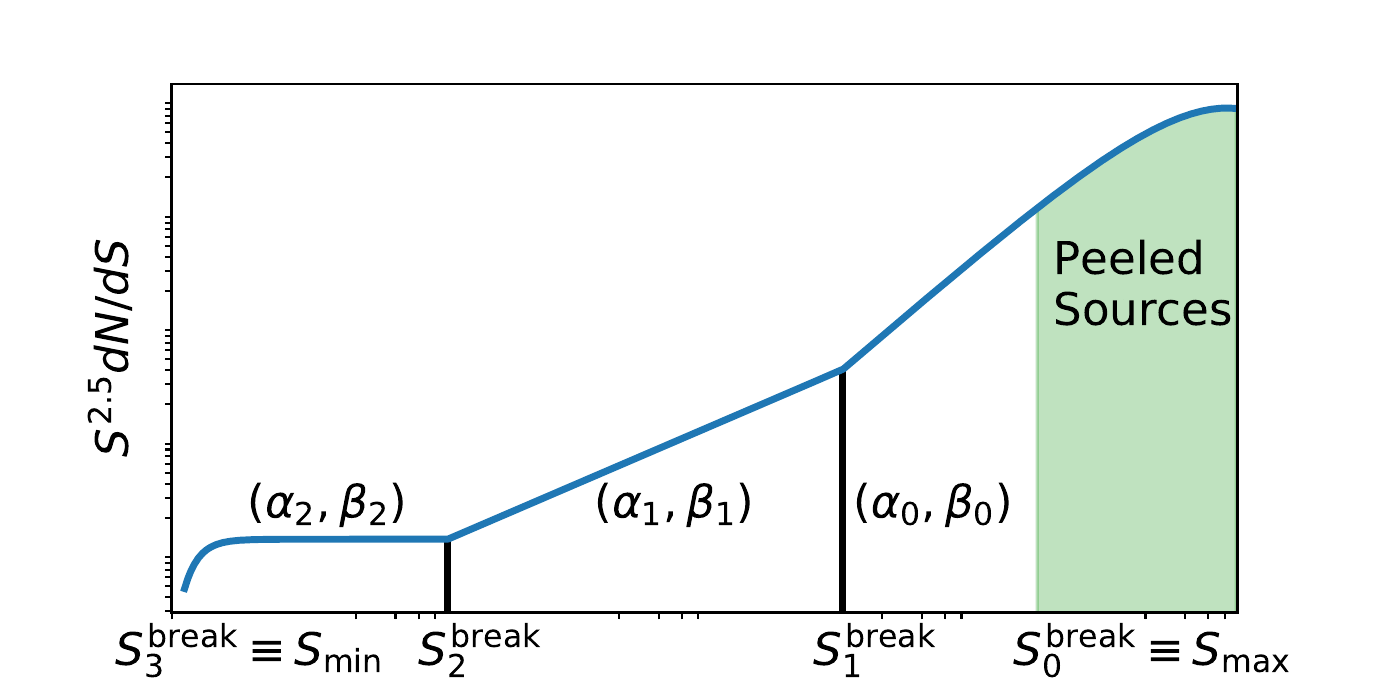}
\caption{Schematic of the broken power-law source count model, with $m=3$.}
\label{fig:source_count_schematic}
\end{figure}

Figure \ref{fig:source_count_schematic} shows a schematic representation of the broken power-law model. The main points of interest are (i) a ``peeled" region at high flux density for which we are reasonably confident of a precise sky model which is subtracted from the visibilities, and for which the lower limit (at $\nu_0$) is $S_{\rm max}$; (ii) there are $m$ regions below this peeling limit; (iii) regions are labelled in order of decreasing flux density, and each is defined by two parameters: the normalisation $\alpha$ and the slope $\beta$, and (iv) an extra ``break" occurs at the physical turnover of the source counts, $S_{\rm min}$.

In practice, we approximate the lower turnover as a sharp cut, and thus the source counts with $m$ regions can be written
\ifx\onecol
\begin{equation}
	\frac{dN}{dS}(\nu_0) = \alpha_i\left(\frac{S}{\rm Jy}\right)^{-\beta_i},\ \  S_{\rm break}^{i+1} \leq S < S_{\rm break}^{i}, \ \ i \in 0,1,...,m-1.
\end{equation}
\else
\begin{align}
\frac{dN}{dS}(\nu_0) &= \alpha_i\left(\frac{S}{\rm Jy}\right)^{-\beta_i}, \\ \nonumber
&S_{\rm break}^{i+1} \leq S < S_{\rm break}^{i}, \ \ i \in 0,1,...,m-1.
\end{align}
\fi
To preserve continuity of the source counts, we define $\alpha_i$ (for $i>0$) as
\begin{equation}
	\alpha_{i} = \alpha_{i-1} S_{{\rm break},i}^{\beta_i - \beta_{i-1}},
\end{equation}
with $\alpha_0 \equiv \alpha$ given by the bright observations.

In this model we find
\begin{equation}
	\label{eq:s2_integral}
	\mu_n = \sum_i \frac{\alpha_i}{n+1-\beta_i}\left[S_{{\rm break},i}^{n+1-\beta_i} - S_{{\rm break},i+1}^{n+1-\beta_i} \right].
\end{equation}

It is not particularly informative to consider the effects of the generalised model on $\vect{C}_{\rm FG}^{\rm Poiss}$, and so we will return to their effects in the next section.

\section{Including Source Clustering}
\label{sec:clustering}
It is well known that extra-galactic sources are clustered, as tracers of the underlying cosmological density field (eg. \citealt{Peebles1971,Percival2001}; and for radio-galaxy examples, see \citealt{Blake2004,Wake2008a}). 
Though this clustering occurs in 3D,  it is projected onto the sky, creating a predictable angular clustering signal. 

It is expected that this clustering will modify the foreground-contributed \HI\ power at scales $k_\perp$ at which the source overdensity power is non-zero.
Thus correctly modelling the effects of cosmological clustering may be an important element of an accurate EoR analysis. 
We note that incorporation of source angular clustering into a foreground covariance model has previously been performed in \cite{Liu2011}, who nevertheless follow a slightly different formalism and restrict themselves to Gaussian correlations.

%\subsection{Derivation of Analytic Model}
We approach this problem in a very general manner, for an arbitrary point-source two-point distribution $P^{ps}(\mathbf{u})$.
%, with $\omega$ the (cosmological) Fourier-dual of the sky-position, $2\pi u$ (i.e. the same quantity as $\vect{u}$ with different Fourier convention).
We note that our comments in \S\ref{sec:review:cov} concerning approximation of the curved $\vect{l}$-space as infinite and Euclidean will also be employed here, as we compare and contrast our improved model to that presented in CHIPS.
Explicitly, we approximate the power spectrum of sources as a function of $u$, without recognising its curved nature. 
As such, our pseudo power spectrum is what one would measure if all sources were expressed in $\vect{l}$ co-ordinates, and these co-ordinates treated as ordinary Euclidean.
This results in a covariance function of sources which is a function of patch separation, expressed in terms of $\vect{l}$.  
Aberrations from this approximation occur only far from beam-centre, for which the contribution to the visibility is negligible. 
A more rigorous derivation in terms of spherical harmonics will be forthcoming in future work.

Our high-level approach is to calculate the covariance of number counts within infinitesimal sky patches due to cosmological clustering, and then apply that to the covariance of the visibilities through a similar formalism as employed in \S\ref{sec:chips:review}.

\subsection{Visibilities with clustered sources}
If a Gaussian field has an over-density power-spectrum $P(u)$, with units ${\rm sr}$, then a realisation of its real-space density field can be evaluated by populating a $u$-space grid with random complex numbers drawn from a standard normal distribution with uniformly-distributed phase, multiplying by the square root of the power spectrum, and then performing an inverse FT \citep{Coles1991}. 
%Explicitly:
%\begin{equation}
%	\label{eq:powerbox}
%	\tilde{\rho}(\vect{x}) = \bar{\rho}\left(\mathcal{F}_{1,1}^{-1}\left[ \sqrt{P(\omega)} \tilde{\mathcal{N}_\omega}e^{i\tilde{X}_\omega} \right] + 1\right),
%\end{equation}
%where $\tilde{\mathcal{N}} \sim {\rm Normal}(0,1)$ and $\tilde{X} \sim {\rm Uniform}[0,2\pi)$, and the alternate Fourier convention is typical of cosmology.
%This is how structured fields are routinely synthetically produced.

Let us for a moment assume that the distribution of flux density on the sky is produced in this manner. 
%dictated purely by Eq. \ref{eq:powerbox}. 
That is, the fluctuations of $\tilde{S}$ are very close to Gaussian (which seems to be cosmologically justified), and that Poisson scatter of individual sources has a negligible effect compared to the clustering itself (we shall amend this assumption soon). Then the field $S(\vect{l})$ is given by
\begin{equation}
	\tilde{S}(\vect{l}) = \bar{S}\left(\mathcal{F}_{0,2\pi}^{-1}\left[ \sqrt{\Omega P(u)} \tilde{\mathcal{N}_u}e^{i\tilde{X}_u} \right] + 1\right)
\end{equation}
where $\Omega$ is the area of integration, $\tilde{\mathcal{N}} \sim {\rm Normal}(0,1)$ and $\tilde{X} \sim {\rm Uniform}[0,2\pi)$.
%, and the alternate Fourier convention is typical of cosmology.

Instead of completely ignoring the Poisson scatter, we may assume that its effects dominate in the second term, but are negligible in the first, in which the other random variables dominate. We will test this assumption in the next section, but here just assert it:
\begin{equation}
	\tilde{S}(\vect{l}) = \left(\mathcal{F}_{0,2\pi}^{-1}\left[ \bar{S} \sqrt{\Omega P(u)} \tilde{\mathcal{N}_u}e^{i\tilde{X}_u} \right] + \int S \widetilde{\frac{dN}{dS}} dS\right).
\end{equation}
Finally, we recall that a visibility is the FT of the beam-attenuated sky brightness with kernel modulated by $f_0$, yielding
\ifx\onecol
\begin{equation}
\label{eq:visibility_clustered}
\tilde{V}(\vect{u},\nu) = \mathcal{F}_{0,2\pi f_0}\left( B(\vect{l},\nu) \left(\mathcal{F}_{0,2\pi}^{-1}\left[ \bar{S}_\nu \sqrt{\Omega P(u)} \tilde{\mathcal{N}_u}e^{i\tilde{X}_u} \right] + B(\vect{l},\nu) \int S \widetilde{\frac{dN_\nu}{dS}} dS\right)\right).
\end{equation}
\else
\begin{align}
\label{eq:visibility_clustered}
\tilde{V}(\vect{u},\nu) = \mathcal{F}_{0,2\pi f_0}\Bigg( B(\vect{l},\nu) \Bigg( & \mathcal{F}_{0,2\pi}^{-1}\left[ \bar{S}_\nu \sqrt{\Omega P(u)} \tilde{\mathcal{N}_u}e^{i\tilde{X}_u} \right] \nonumber \\
 &+\left.\left. B(\vect{l},\nu) \int S \widetilde{\frac{dN_\nu}{dS}} dS\right)\right).
\end{align}
\fi

\subsection{Statistics of visibilities}
Using Eq. \ref{eq:visibility_clustered} as the basis for a stochastic visibility, we may proceed to derive the statistics of the visibility -- specifically its expectation and covariance between frequencies.

While the formula for the visibility has grown considerably more complex as compared to Eq. \ref{eq:visibility}, the expected visibility remains the same -- that is, the FT of the beam normalised by mean flux density -- because the first term in Eq. \ref{eq:visibility_clustered} has a mean of zero (due to the $\N$). This is intuitively understood because an average over realisations of clustered foregrounds will change phase uniformly, and therefore average to zero. 

Our primary goal is the covariance of the visibility. We first note that the second term in Eq. \ref{eq:visibility_clustered} is precisely the same as the uniform-sky visibility, Eq. \ref{eq:visibility}, so that clustering is an additive modification. Furthermore, covariance between the first and last term is zero, since the random part of each is independent (in fact, even if we had not used the deterministic $\bar{S}$ in the first term, and had rather kept it as the more general random variable found in the second term, the covariance is still zero). Thus we have
\begin{equation}
	\label{eq:c_as_sum}
	\vect{C}_{\rm FG} = \vect{C}_{\rm FG}^{\rm Clust} + \vect{C}_{\rm FG}^{\rm Poiss},
\end{equation}
where 
\begin{align}
	 \vect{C}_{\rm FG}^{\rm Clust} = {\rm Cov} & \left[\mathcal{F}_{0,2\pi f'_0}\left( \bar{S}' B(\vect{l},\nu') \mathcal{F}_{0,2\pi}^{-1}\left[  \sqrt{P(u)} \tilde{\mathcal{N}_u}e^{i\tilde{X}_u} \right]\right) \right. , \nonumber \\
	 & \left. \mathcal{F}_{0,2\pi f''_0}\left(\bar{S}'' B(\vect{l},\nu'') \mathcal{F}_{0,2\pi}^{-1}\left[  \sqrt{\Omega P(u)} \tilde{\mathcal{N}_u}e^{i\tilde{X}_u} \right]\right) \right].
\end{align}
Now, the outer FT can be pushed through, by using the convolution theorem, and the fact that $\mathcal{F}_{0,2\pi a} f(l) = (1/a) \hat{f}(l/a)$:
\ifx\onecol
\begin{align}
 \vect{C}_{\rm FG}^{\rm Clust} = \frac{\bar{S}^2}{(f'_0f''_0)^{1+\gamma}} {\rm Cov} & \left[\left(  \hat{B'}(\vect{l}/f'_0) \otimes  \left[  \sqrt{\Omega P(f'_0 u)} \N_{f'_0 u}e^{i\tilde{X}_{f'_0 u}} \right]\right) \right. , \nonumber \\
& \left. \left(  \hat{B''}(\vect{l}/f''_0) \otimes \left[  \sqrt{\Omega P(f''_0 u)} \N_{f''_0u}e^{i\tilde{X}_{f''_0 u}} \right]\right) \right].
\end{align}
\else
\begin{align}
\vect{C}_{\rm FG}^{\rm Clust} = &\frac{\bar{S}^2}{(f'_0f''_0)^{1+\gamma}} \times  \nonumber \\
&{\rm Cov} \left[\left(  \hat{B'}(\vect{l}/f'_0) \otimes  \left[  \sqrt{\Omega P(f'_0 u)} \N_{f'_0 u}e^{i\tilde{X}_{f'_0 u}} \right]\right) \right. , \nonumber \\
& \ \ \ \left. \left(  \hat{B''}(\vect{l}/f''_0) \otimes \left[  \sqrt{\Omega P(f''_0 u)} \N_{f''_0 u}e^{i\tilde{X}_{f''_0 u}} \right]\right) \right].
\end{align}
\fi
The convolution is again just a sum, so the covariance is the sum of covariance of all pairs (we also make the identification that $\bar{S} \equiv \mu_1$):
\ifx\onecol
\begin{align}
 \vect{C}_{\rm FG}^{\rm Clust} = \frac{\mu_1^2}{(f'_0f''_0)^{1+\gamma}} \int \int  & \widehat{B'(\vect{l}/f'_0)} (\vect{u}-\vect{u}_1)  \widehat{B''(\vect{l}/f''_0)}(\vect{u}-\vect{u}_2) \Omega \sqrt{P( f'_0 u_1)}  \sqrt{\P(f''_0 u_2)} \nonumber \\  
& \times {\rm Cov} \left[\N_{f'_0 u_1}e^{i\tilde{X}_{f'_0 u_1}} ,   \N_{f''_0u_2}e^{i\tilde{X}_{f''_0u_2}} \right] d\vect{u}_1 d\vect{u}_2.
\end{align}
\else
\begin{align}
\vect{C}_{\rm FG}^{\rm Clust} = &\frac{\mu_1^2}{(f'_0f''_0)^{1+\gamma}}\times \nonumber \\
 \int \int  & \widehat{B'(\vect{l}/f'_0)} (\vect{u}-\vect{u}_1)  \widehat{B''(\vect{l}/f''_0)}(\vect{u}-\vect{u}_2) \times \nonumber \\
 &\Omega \sqrt{P(f'_0 u_1)}  \sqrt{P( f''_0 u_2)} \times \nonumber \\  
& {\rm Cov} \left[\N_{f'_0 u_1}e^{i\tilde{X}_{f'_0 u_1}} ,   \N_{f''_0 u_2}e^{i\tilde{X}_{f''_0 u_2}} \right] d^2\vect{u}_1 d^2\vect{u}_2.
\end{align}
\fi
Since each draw of $\N$ is independent, the covariance term remaining here must be $\delta(f'_0 u_1 - f''_0 u_2)/\Omega$, where the solid angle factor ensures that the covariance is dimensionless as it ought to be. So, similarly to the Poisson case, the double-integral collapses to a single integral:
\ifx\onecol
\begin{equation}
	\label{eq:general_clustering_covariance}
 \vect{C}_{\rm FG}^{\rm Clust} = \frac{\mu_1^2}{(f'_0f''_0)^{1+\gamma}}\frac{f'_0}{f''_0} \int \widehat{B'(\vect{l}/f'_0)} (\vect{u}-\vect{u}_1)  \widehat{B''(\vect{l}/f''_0)}(\vect{u}-f'_0/f''_0\vect{u}_1) \Omega P(f'_0 u_1) d^2\vect{u}_1.
\end{equation}
\else
\begin{align}
\label{eq:general_clustering_covariance}
\vect{C}_{\rm FG}^{\rm Clust} = & \frac{\mu_1^2 p}{(f'_0f''_0)^{1+\gamma}} \int P(f'_0 u_1) \times \nonumber \\
&\ \ \ \widehat{B'(\vect{l}/f'_0)} (\vect{u}-\vect{u}_1)  \widehat{B''(\vect{l}/f''_0)}(\vect{u}-p\vect{u}_1) d^2\vect{u}_1,
\end{align}
\fi
where we have set $p = f'_0/f''_0$.
This is the most general form for the clustering covariance term that can be derived. 

A few remarks are appropriate. Firstly, the overall source-count-based normalisation is different ($\mu_1^2$) in this term than in $\vect{C}_{\rm FG}^{\rm Poiss}$. 
Secondly, there is again a complex interaction between frequencies and $\vect{u}$, which will bring more structure into the 2D PS than the purely Poisson covariance. The form of Eq. \ref{eq:general_clustering_covariance} is rather complicated, but for the variance, it simplifies into a pure convolution between the square beam and the power spectrum. 

Thus in general, the full covariance, in the presence of source clustering, can be evaluated using a 2-dimensional integration.

\subsection{Model Simplifications}
While Eq. \ref{eq:general_clustering_covariance} is the most general form of the clustering covariance, if we assume a simple model for the beam (i.e. a frequency-dependent Gaussian beam), some simplifications can be made. 

The frequency-dependence of the Gaussian beam arises through the width of the beam:
\begin{equation}
	\sigma \simeq \epsilon c/ \nu D,
\end{equation}
where $\epsilon \simeq 0.42$ is the scaling from an Airy disk to a Gaussian width and $D$ is the tile diameter. Because of this simple relationship, it turns out that the FT of the frequency-dependent beam where the argument is $\vect{l}/\nu$ collapses into the standard FT of the beam at $\nu_{\rm low}$, so we have
\ifx\onecol
\begin{equation}
	\label{eq:circular_clustering_covariance}
	\vect{C}_{\rm FG}^{\rm Clust} = \frac{\mu_1^2p}{(f'_0f''_0)^{1+\gamma}}\int \hat{B_0}(\vect{u}-\vect{u}_1)  \hat{B_0}(\vect{u}-p\vect{u}_1) \Omega P(f'_0 u_1) \frac{d^2\vect{u}_1}{\rm sr}.
\end{equation}
\else
\begin{equation}
\label{eq:circular_clustering_covariance}
\vect{C}_{\rm FG}^{\rm Clust} = \frac{\mu_1^2p}{(f'_0f''_0)^{1+\gamma}}\int  P( f'_0 u_1)  \hat{B_0}(\vect{u}-\vect{u}_1)  \hat{B_0}(\vect{u}-p\vect{u}_1)d^2\vect{u}_1.
\end{equation}
\fi
The FT of the beam is $\hat{B} = 2\pi \sigma^2 \exp(-2\pi^2\sigma^2 u^2)$, so inserting this, setting $y=2\pi\sigma^2$, and $p=f'_0/f''_0$, and expanding the exponents, we have
\ifx\onecol
\begin{equation}
	\vect{C}_{\rm FG}^{\rm Clust} = \frac{\mu_1^2}{(f'_0f''_0)^{1+\gamma}}p \int y^2 e^{-\pi y[2u^2 + u_1^2(1+p^2) - 2\vect{u}\cdot\vect{u}_1(1+p)]} P( f'_0 u_1) d^2\vect{u}_1.
\end{equation}
\else
\begin{align}
\vect{C}_{\rm FG}^{\rm Clust} = & \frac{\mu_1^2p}{(f'_0f''_0)^{1+\gamma}} \int P( f'_0 u_1) \times \nonumber \\
&y^2 e^{-\pi y[2u^2 + u_1^2(1+p^2) - 2\vect{u}\cdot\vect{u}_1(1+p)]} d^2\vect{u}_1.
\end{align}
\fi
Acknowledging that the solution is circularly symmetric, we assume that $v=0$, and convert to polar co-ordinates so that $\vect{u}\cdot\vect{u}_1 = uu_1 \cos\theta$:
\ifx\onecol
\begin{equation}
\vect{C}_{\rm FG}^{\rm Clust} = \frac{\mu_1^2 p y^2}{(f'_0f''_0)^{1+\gamma}} e^{-2\pi y u^2} \int e^{-\pi yu_1^2(1+p^2)} \int e^{2\pi y uu_1\cos\theta (1+p)} P( f'_0 u_1)u_1 d\theta du_1 .
\end{equation}
\else
\begin{align}
\vect{C}_{\rm FG}^{\rm Clust} = \frac{\mu_1^2 y^2}{(f'_0f''_0)^{1+\gamma}}p &e^{-2\pi y u^2} \int e^{-\pi yu_1^2(1+p^2)} P(f'_0 u_1) \times \nonumber \\
&\int e^{2\pi y uu_1\cos\theta (1+p)}u_1 d\theta du_1 .
\end{align}
\fi
Integrating over $\theta$ yields
\ifx\onecol
\begin{equation}
\vect{C}_{\rm FG}^{\rm Clust} = \frac{2\pi \mu_1^2 py^2}{(f'_0f''_0)^{1+\gamma}} e^{-2\pi y u^2} \int e^{-\pi yu_1^2(1+p^2)} I_0(2\pi y uu_1(1+p)) P( f'_0 u_1)u_1 du_1,
\end{equation}
\else
\begin{align}
\vect{C}_{\rm FG}^{\rm Clust} = \frac{2\pi \mu_1^2 p y^2}{(f'_0f''_0)^{1+\gamma}} &e^{-2\pi y u^2}  \int  e^{-\pi yu_1^2(1+p^2)}   P( f'_0 u_1) \times \nonumber \\
& I_0(2\pi y uu_1(1+p))u_1du_1,
\end{align}
\fi
where $I_0$ is the zeroth-order modified Bessel function of the first kind. Thus the visibility covariance for any isotropic point-source PS with a Gaussian beam reduces to a single integral. 

Let us further assume that the point-source PS is a power-law:
\begin{equation}
	\label{eq:power_law_ps}
	\frac{P(u)}{\rm sr} = \left(\frac{u}{u_0}\right)^{-\kappa}.
\end{equation}
In this case, the Bessel function may be expanded using its power-series:
\begin{equation}
	I_0(z) = \sum_{k=0}^\infty \frac{(z/2)^{2k}}{k!^2},
\end{equation}
and each term may be analytically integrated to yield Euler gamma functions. The resulting sum may be simplified to 
\ifx\onecol
\begin{equation}
	\label{eq:clustering_cov_gauss_pl}
	\vect{C}_{\rm Clust} =	(\pi y)^{\kappa/2} y \mu_1^2 u_0^\kappa \frac{(1+p^2)^{-a}}{{f'}_0^{\kappa+\gamma}{f''}_0^{2+\gamma}}   \left[ \Gamma(a) e^{-2\pi y u^2} {}_1F_1\left(a; 1; \pi y \frac{(1+p)^2}{1+p^2} u^2 \right)\right],
\end{equation}
\else
\begin{align}
\label{eq:clustering_cov_gauss_pl}
\vect{C}_{\rm Clust} =	&(\pi y)^{\kappa/2} y \mu_1^2 u_0^\kappa \frac{(1+p^2)^{-a}}{{f'}_0^{\kappa+\gamma}{f''}_0^{2+\gamma}}  \times \nonumber \\
& \left[ \Gamma(a) e^{-2\pi y u^2} {}_1F_1\left(a; 1; \pi y \frac{(1+p)^2}{1+p^2} u^2 \right)\right],
\end{align}
\fi
% \frac{p {f'}_0^{-\kappa} (1+p^2)^{-a}}{(f'_0f''_0)^{1+\gamma}}
where $y=2\pi\sigma^2$, and ${}_1F_1$ is the Kummer (confluent) hypergeometric function, and $a = 1 - \kappa/2$.
At large $u$, Kummer's function asymptotes to
\begin{equation}
	{}_1F_1(a;b;x) = e^x x^{a-b} \Gamma(b)/\Gamma(a),
\end{equation}
\ifx\comments
	{\color{red}
	so that the last factor in square brackets becomes
	\ifx\onecol
	\begin{equation}
		\label{eq:large_u_approx}
		 \Gamma(a) e^{-2\pi y u^2} {}_1F_1\left(a; 1; \pi y \frac{(1+p)^2}{1+p^2} u^2 \right) = e^{\pi y u^2 (q -2)} (\pi y q)^{-\kappa/2} u^{-\kappa},
	\end{equation}
	\else
	\begin{equation}
	\label{eq:large_u_approx}
	e^{\pi y u^2 (q -2)} (\pi y q)^{-\kappa/2} u^{-\kappa},
	\end{equation}
	\fi
	where $q = (1+p)^2/(1+p^2)$ has a maximum of 2 when $f'_0 \equiv f''_0$.}
\fi
so that we can approximate the covariance (under the assumption of a Gaussian beam and power-law source spectrum) as
\begin{equation}
	\label{eq:large_u_approx}
	\cclust = y\mu_1^2 Q_\nu P(u) e^{\pi y u^2 (q-2)},
\end{equation}
where $q = (1+p)^2/(1+p^2)$ has a maximum of 2 when $f'_0 \equiv f''_0$, and
\begin{equation}
	Q_\nu = \frac{q^{-\kappa/2}(1+p^2)^{\kappa/2-1}}{{f'}_0^{\kappa+\gamma}{f''}_0^{2+\gamma}}.
\end{equation}
Thus, the variance on intermediate-to-small scales collapses to a single power-law with slope $-\kappa$, and otherwise is a decaying exponential, modified by the same power-law. 

\begin{figure}
\centering
\includegraphics[width=\linewidth,trim=0cm 0cm 1cm 0cm]{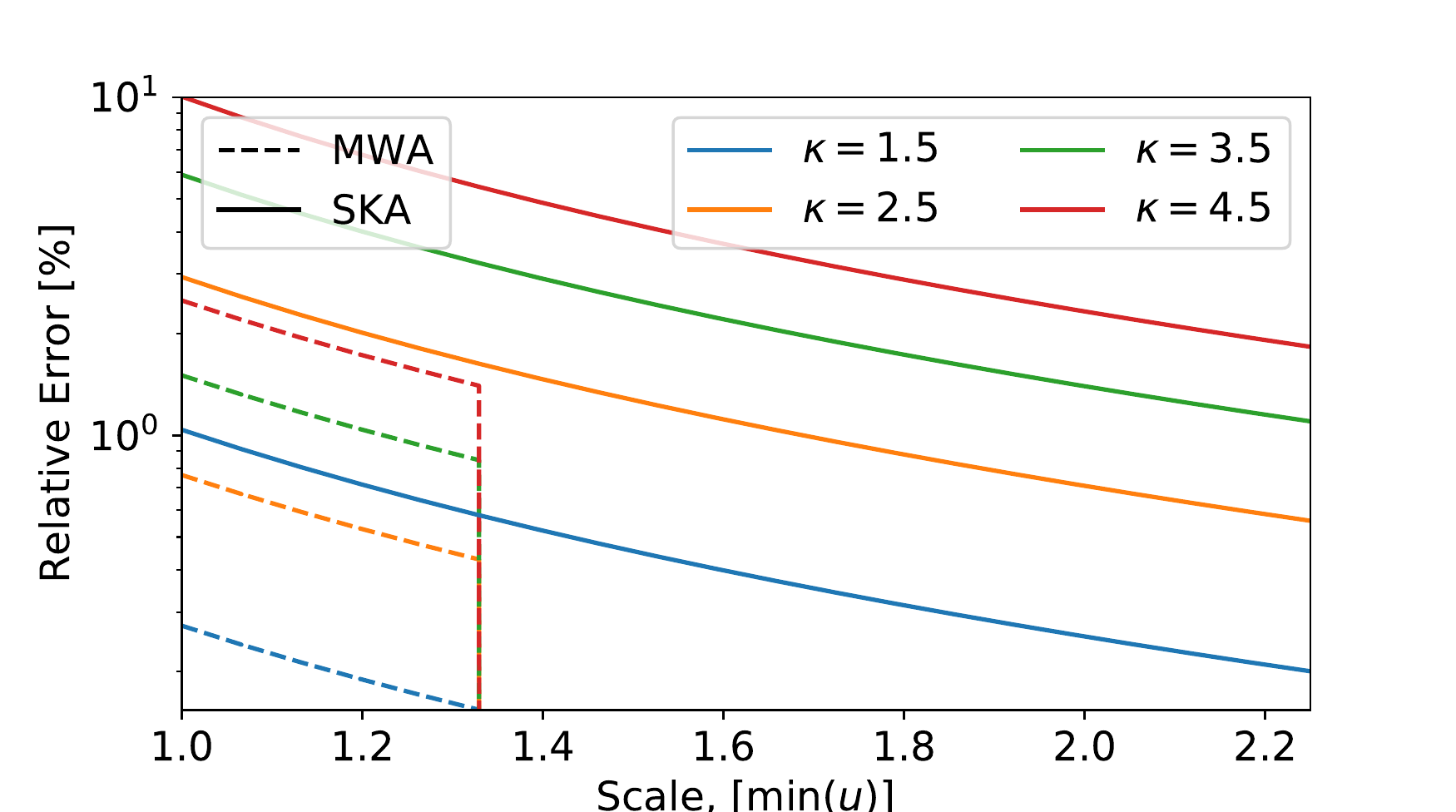}
\caption{Error induced by the large-$u$ approximation (cf. Eq. \ref{eq:large_u_approx}) as a function of scale for various values of the power-spectral slope $\kappa$. The $x$-axis units are multiples of the smallest measured scale (i.e. the shortest baseline of the array at 150 MHz). This scale changes for different instruments, and we show the MWA and proposed SKA configurations here. The SKA has larger error on its smallest baseline, predominantly owing to its smaller beam size. Regardless, the approximation holds to within 10\% over all measured baselines in realistic experiments. The truncation in the MWA curves is due to numerical precision in calculating the LHS of Eq. \ref{eq:large_u_approx}.}
\label{fig:approximation_error}
\end{figure}

It is helpful to determine the range of scales on which this large-$u$ approximation holds. Figure \ref{fig:approximation_error} shows the relative error induced by the approximation over a range of scales, and for various values of the power-spectral slope. The $x$-axis units are multiples of the smallest measured scale, i.e. the shortest baseline of the array at 150 MHz.
To assess realistic instruments, we use the (128-tile) configuration of the MWA, and the proposed configuration of SKA1-LOW \citep{Dewdney2016}. While the SKA shows an increase in the induced error, primarily due to the reduced beam width, it is clear that over a large range of physical models of the point-source power spectrum, the maximum expected error is less than 10\%. 
In the numerical implementations used for this paper, the large-$u$ approximation is employed, as it is more stable for the majority of scales considered. 
%In the numerical implementations used for this paper, the full solution is used, up until the point at which the large-$u$ approximation is more numerically stable.
%However, it is useful to bear in mind that the large-$u$ approximation is typically valid over the whole observation, which aids intuition.

\subsection{Summary of Clustered Covariance Model}
\label{sec:summary_params}
Several iterations of the total analytic data covariance of the point-source foregrounds have been presented thus far, under a range of simplifying assumptions.
This presentation allows for the continued development of the covariance formalism by relaxation of these assumptions. 
However, for the remainder of this introductory paper, we shall limit ourselves to a definite simple prescription, so as to enable an initial survey of the effects of its parameters.

The total data covariance is approximated by a sum of a Poisson and Clustering term (cf. Eq. \ref{eq:c_as_sum}) -- an approximation which is validated in the next subsection.
Each term is analytically derived using a set of empirical models of the observed sky -- namely (i) the distribution of spectral slopes of sources; (ii) the distribution of flux density of sources at a reference frequency $\nu_0$; (iii) the telescope's spatial attenuation pattern (or the beam); and
(iv) the spatial distribution of sources on the sky, parameterised by its isotropic power spectrum.

Henceforth we exclusively employ the following models for these four components:
(i) a single universal spectral slope for all sources: $\gamma=0.8$;
(ii) an arbitrarily broken power-law with hard minimum $S_{\rm min}$ and peeling limit $S_{\rm max}$ (cf. \S\ref{sec:source_counts});
(iii) a frequency-dependent Gaussian beam, $\exp(-l^2/2\sigma_\nu^2)$, with $\sigma = 0.42c/\nu D$, and $D$ the effective tile diameter; and
(iv) a single-power law power spectrum, according to Eq. \ref{eq:power_law_ps}.
Under these assumptions, the total covariance is given by Eqs. \ref{eq:c_as_sum}, \ref{eq:poisson_gaussian}, \ref{eq:clustering_cov_gauss_pl} and \ref{eq:mu} (bearing in mind the approximation of Eq. \ref{eq:large_u_approx}).

For ease of reference, we provide a summary of all model parameters in Table \ref{tab:parameters}. In this table, values labelled ``Fixed" are considered to be well-known (potentially per telescope) so that they may be kept constant throughout the rest of the analysis (in which case their values are given by the table). The rest of the parameters are varied to assess their impact on results -- either in toy models (cf. next subsection) or in the context of a synthetic EoR observation (cf. \S\ref{sec:application}). In all cases, the ``fiducial model" will correspond to Table \ref{tab:parameters}.

\begin{deluxetable*}{llll}
	\tabletypesize{\footnotesize} 
	\tablecolumns{4} 
	\tablewidth{0pt} 
	\tablecaption{Summary of analytic covariance model parameters.}
	\tablehead{\colhead{Param.} & Description & \colhead{(Fiducial) Value} & Notes} 
	\startdata
	$\gamma$ & Single universal spectral slope of sources at $\nu_0$ & 0.8  & Fixed \\
	$S_{\rm min}$ & Minimum flux density of sources at $\nu_0$ & $10^{-1}$ mJy & Varied \\
	$S_{\rm max}$ & Peeling limit at $\nu_0$ & $\{30,1\}$ mJy & Varied \\
	$S_{{\rm break},i+1}$ & Position of break in broken power-law source counts between $i^{th}$ and $(i+1)^{th}$ region & [6] mJy & Fixed \\ 
	$\alpha$ & Normalisation of source count distribution in highest flux density region & 6998 ${\rm Jy^{-1} {\rm sr}^{-1}}$ & Fixed \\
	$\beta_i$ & Logarithmic slope of source counts in $i^{th}$ region & [1.54, 1.95] & Varied \\  
	$D$ & Effective tile diameter of telescope & $\{4,35\} $ m &  Fixed \\
	$u_0$ & Normalisation of point-source power spectrum & 0.05 & Varied \\
	$\kappa$ & Slope of point-source power spectrum & 1.5 & Varied 
	\enddata
	\tablecomments{A summary of all parameters in the analytic covariance model, under the simplified models outlined in \S\ref{sec:summary_params}. Curly braces in the value column indicate that the parameter is dependent on the telescope itself, and the values given are the fiducial values for an MWA-like and SKA-like array respectively. Square brackets indicate that the parameter is (potentially) a vector of values, i.e. there is a value for the parameter in each subregion of the source count distribution. In this case, the regions are taken from highest to lowest flux density. }
	\label{tab:parameters}
\end{deluxetable*}

\subsection{Model Features in the Covariance}
The analytic covariance follows a straightforward form -- at large scales, the clustering term dominates, while on small scales, the Poisson term dominates. 
The simplest way to determine the effects of the model parameters is to determine the scale $u_\star$ at which these two terms are equal -- the smaller the scale, the more important the clustering term becomes. 

The general solution is rather intractable, however inspection of Fig. \ref{fig:compare_clustering_only} reveals three key observations: (i) the covariance between different frequencies is always subdominant to the variance
at $u_\star$ (for either the covariance or variance), due to the exponential cut-off, (ii) scales within a physically reasonable observation are typically well-approximated by the large-$u$ approximation (cf. Eq. \ref{eq:large_u_approx}), and (iii) the transition scale is similar for the covariance between frequencies and variance.

%we make the following three observations: (i) the most important part of the covariance is the variance (which dominates the resulting power spectrum); (ii) wherever the clustering term dominates in the variance, it seems also to dominate in the covariance, and (iii) the amplitude of the covariance is always negligible compared to the variance on scales at which the Poisson term dominates.

These three observations lead us to consider just the variance itself as a measure of the effects of the parameters. By equating Eq. \ref{eq:large_u_approx} with Eq. \ref{eq:poisson_gaussian} for $\nu'=\nu''$, we find that 
\begin{equation}
\label{eq:ustar}
u_\star = \left(\frac{\mu_1^2}{\mu_2}\right)^{1/\kappa}  u_0 .
\end{equation}

\begin{figure*}
	\centering
	\includegraphics[width=0.85\linewidth]{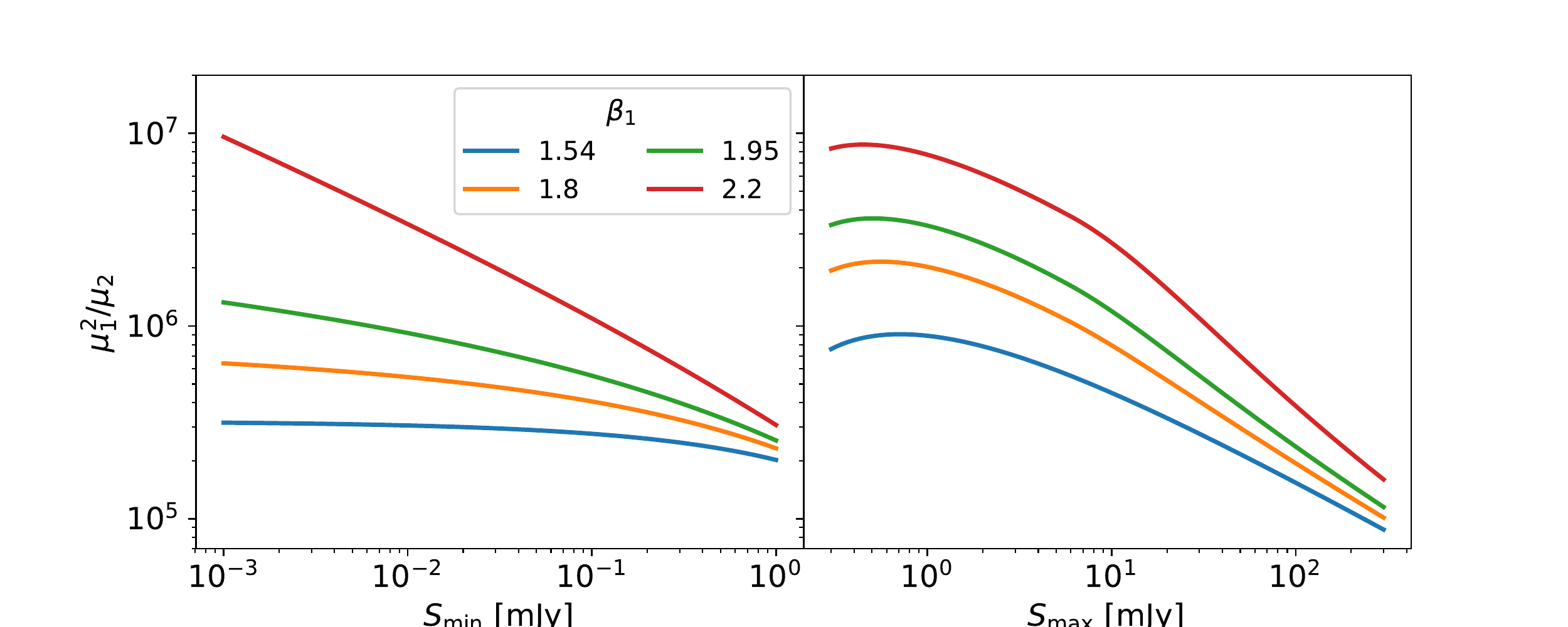}
	\caption{Values of the ratio of moments of the source count distribution, as presented in \ref{eq:ustar}. The $x$-axes show values of $S_{\rm min}$ and $S_{\rm max}$ (left and right panels respectively), while the $y$-axis shows the ratio of the moments. Different colours indicate different values of the power-law slope in the star-forming-galaxy-dominated region of the source counts. The ratio tends to favour a strong clustering signal for lower values of both bounding flux densities, as well as steeper slopes. As expected, the steepness of the slope is not as effective when either boundary is moved higher. }
	\label{fig:source_count_comparison}
\end{figure*}

We first consider the relationships between parameters. Clearly, the amplitude of the point-source power spectrum has a linear effect on the importance of the clustering term.
The more interesting relationship is that of the ratio of moments of the source count distribution, which involves 5 out of the 9 parameters of the model ($S_{\rm min}$, $S_{\rm max}$, $\alpha$, $\beta_i$ and $\sbrk$).
In Fig. \ref{fig:source_count_comparison} we show the effects of modifying a selection of these parameters on the ratio. The parameters varied here are the flux-density boundaries (one of which is entirely unknown, and the other dependent on survey design) and the slope of the source counts in the star-forming-galaxy-dominated region of the source counts.
The position of the break is kept at its default value, and a (small) change in this value is not expected to modify the qualitative behaviour of the plot. Furthermore, the slope and normalisation of the source counts above the break are taken as relatively well known, and therefore fixed. 

The left and right panels of Fig. \ref{fig:source_count_comparison} respectively show the ratio as a function of $S_{\rm min}$ and $S_{\rm max}$, with different coloured lines indicating various source count slopes. Since a higher ratio corresponds to a greater $u_\star$ (and thus a greater contribution from the clustering signal), we find that decreasing either boundary results in an increase in the clustering contribution. This increase is more significant for steeper slopes, $\beta_1$. When $S_{\rm max}$ is below the break (which may be the case for the SKA), there is a turnover, and lower peeling limits correspond to a reduction in the fractional contribution of clustering. 

Along with the relationships between parameters, we would also like to consider the absolute value of $u_\star$ under reasonable parameter choices. This is significant, because a very low $u_\star$ might be below the limit of measurement for a given instrument, which has a physical minimum baseline length. 
To do this, we take two extreme cases, under the assumption of MWA-like and SKA-like parameters -- ($S_{\rm min}$,$S_{\rm max}$, $\beta_1$, $u_0$, $\kappa$) = (0.1 mJy, 30 mJy, 1.54, 1, 4.5) and (0.001 mJy, 1 mJy, 2.2, 0.01, 1.5). For these two sets of parameters, we find $x_\star = u_\star \lambda_{150} \approx (32, 6\times10^3)$ m.
The first of these is below the minimum baseline for current SKA specifications, while the latter implies that clustering dominates over all relevant baselines for both instruments.

To move forward, we require a more robust physical model of both the source counts in the very-faint regime, and the spatial distribution of sources. We do not pursue these further in this work, but note that once these have been established, a firm estimate of the relative importance of the clustering term can be attained given the instrumental specifications.

\subsection{Comparison to Toy Simulations}
Before moving on to realistic simulations, we do well to gain intuition from toy models in which we can make as many simplifications as possible. 
To do this, we generate a distribution of sources on the ``sky'' (or a finite Euclidean space which we use as an approximation to $\vect{l}$) consistent with an underlying power-law source-count distribution with Poisson scatter and a power-law power spectrum, using the \textsc{powerbox} software\footnote{Found at https://github.com/steven-murray/powerbox}. 
We then bin these sources separately at each frequency, using different cell sizes so that the corresponding DFT at each frequency contains the correct values of $\vect{u}$.
Finally, we attenuate the resulting sky with a Gaussian beam before taking the 2D DFT to calculate visibilities. 
Each visibility is circularly averaged (its real and imaginary parts, as well as its square) to form a 1D visibility, and the final covariance is calculated as the mean over many realisations of the squared visibility minus the mean of the visibilities squared.

A physical power spectrum is expected to turnover at some scale, and fall to zero at $\omega=0$. However, as long as this scale is within the resolution of our simulation, it will be completely undetectable (in fact, the finite resolution, $d$, of the simulation imposes a hard cut on the power spectrum at roughly $\omega = 2\pi/d$). Thus we only trust the results well above this limit (and conversely, below the scale limit imposed by the finite box size of the simulation).

\subsubsection{Basic Tests}
\begin{figure*}
\centering
\includegraphics[width=\linewidth, trim=1cm 0cm 2cm 0cm]{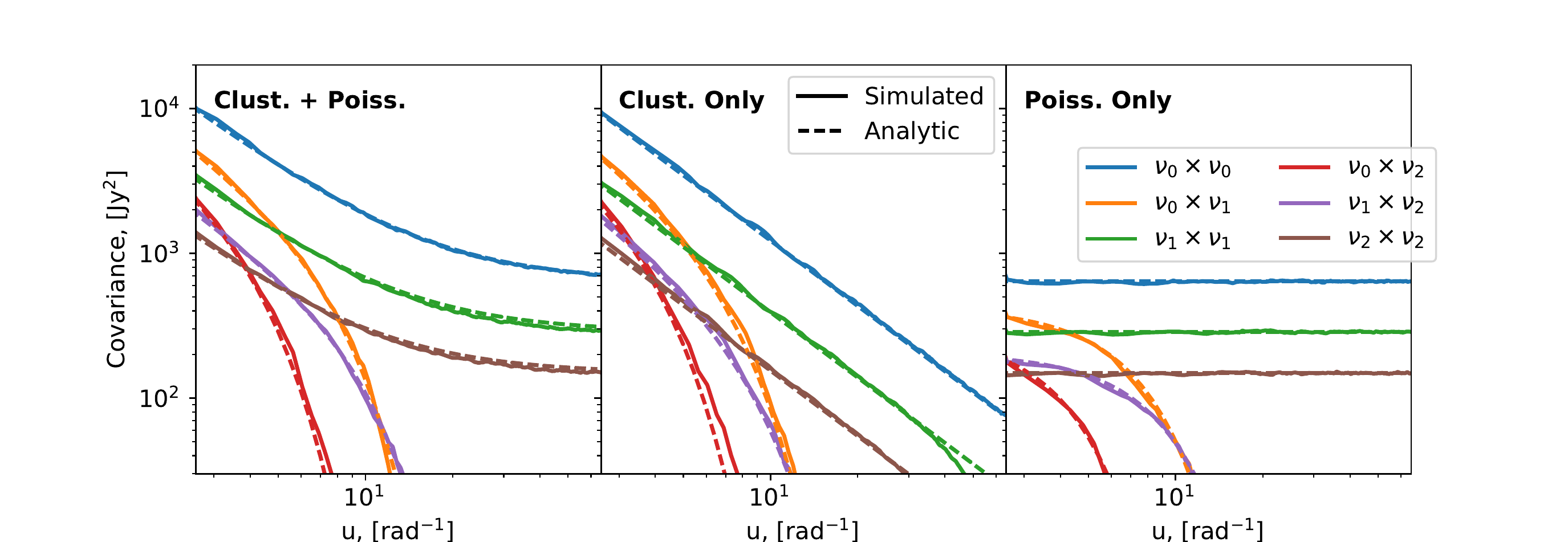}
\caption{Covariance between frequency channels of visibilities including only point sources. Only two frequency channels are shown for clarity, and these are exaggerated. The reference frequency $\nu_0$ = 150.0MHz, and the displayed frequencies are $f_0 = (1.0,1.25, 1.5)$. The box-size of the simulation is four times the FWHM of the beam, at 1.68, with 256 grid cells per side, and 200 realisations were used in order to generate the covariance. The power-law power spectrum has $(u_0,\kappa)=(0.02,1.5)$. Blue, green and brown lines indicate variances at each frequency as a function of the perpendicular scale $u$, while the orange, purple and maroon lines indicate covariances between the frequencies. Solid lines show fully simulated results, while dashed lines show the result of Eq. \ref{eq:general_clustering_covariance}. The LHS panel shows the results of the full covariance, with both clustering and Poisson scatter present, while the centre and RHS panels show the results with just clustering and Poisson terms respectively. Eq. \ref{eq:general_clustering_covariance} assumes these stochastic effects can be separated, and the LHS panel confirms the validity of this assumption, revealing no significant departure of the simulation from the model. The primary effect of including Poisson scatter is to add a constant to the variances, to which the solutions asymptote.
}
\label{fig:compare_clustering_only}
\end{figure*}

Our first task is to ensure that the model, Eq. \ref{eq:clustering_cov_gauss_pl}, provides a good description of the simulated data, and to examine its features. 
We show the results of a comparison of the simulated covariance to the model in Fig. \ref{fig:compare_clustering_only}. In this section, as we are comparing to (relatively) expensive simulations and we are interested only in the fidelity of the comparison, rather than the absolute value of the result, we use modified source count limits: $(S_{\rm min},S_{\rm max}) = (0.1,1)$ Jy. Apart from this modification, we use the default model from Table \ref{tab:parameters}, with $(u_0,\kappa) = (0.02,1.5)$.

The immediate comparison is between solid and dashed lines, which indicate simulation and model respectively. These show very good agreement across the range of scales $u$ shown here. The range shown in the plot corresponds roughly to the resolution limits of the simulations. 

Of particular interest in Fig. \ref{fig:compare_clustering_only} is the comparison of model to simulation in the presence of Poisson scatter. 
Recall that the derivation of the model incorporated the assumption that the stochasticity of clustering and Poisson scatter could be separated into two terms.
The left-hand panel shows the results of the model including this assumption against the simulations, which require no such assumption. 
Though there is potentially some small departure at low $u$, the overall effect is negligible. 
The middle panel shows a comparison in which no Poisson scatter is used at all, which we expect to be in tighter agreement. 
The improvement however is almost visually imperceptible. 
Thus we conclude that the assumption of separability is indeed valid.

We finally note the features of the model. Clearly, $\vect{C}_{\rm FG}^{\rm Poiss}$ (right-hand panel) has a constant variance, with exponential decay in the covariance (as made clear in Eq. \ref{eq:poisson_gaussian_covariance}). 
This chromaticity of the instrument is what results in the so-called foreground `wedge` in the 2D PS.
The clustering term, $\vect{C}_{\rm FG}^{\rm Clust}$ displays a power-law for the variance, of the same slope as the point-source PS, and normalisation dictated by a combination of factors, including the normalisation of the point-source PS. 
The covariance is also a power-law at low-$u$, but experiences exponential decay at mid-$u$. 
We expect this exponential decay to be a feature of the \textit{beam}, rather than the PS, while the low-$u$ slope is clearly dominated by the latter.
Their combination in the left-hand panel is very close to a pure addition of these terms.
The resulting variance is dominated by the clustering term on large scales, and the Poisson term on small scales. 
The transition scale will be dictated by the relative normalisation between the two, along with the form of the point-source PS.
A similar trend emerges for the covariance, with large scales dominated by $\vect{C}_{\rm FG}^{\rm Clust}$, and small scales behaving similarly between the two terms.
Thus in summary we expect that  $\vect{C}_{\rm FG}^{\rm Clust}$ may be important (dependent on its normalisation) for describing the covariance relevant for EoR analyses.

\begin{figure*}
\centering
\includegraphics[width=0.9\linewidth]{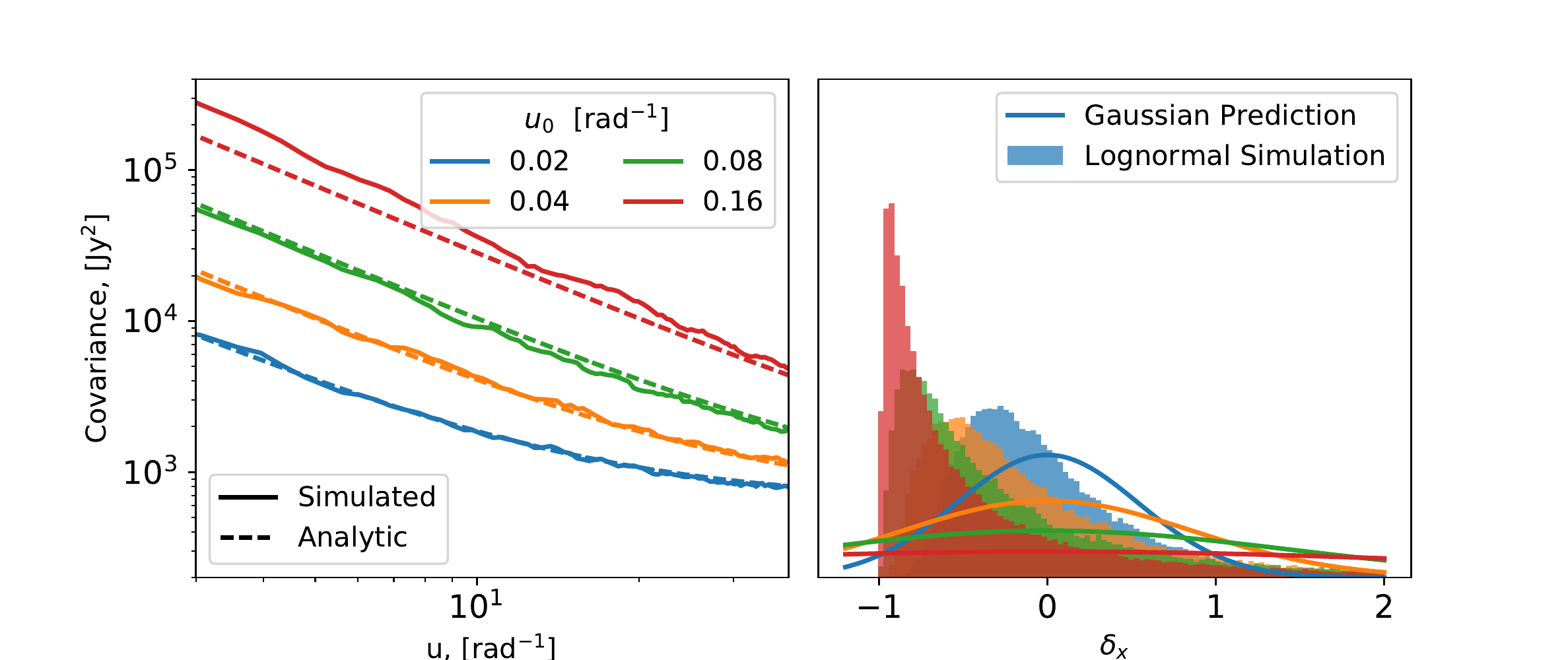}
\caption{Comparison of model covariance to simulations with varying magnitudes of bin-to-bin variance, $\sigma^2$. The point-source PS has $\kappa=$1.5, with varying $\omega_0$, as shown by line colour. Model predictions (shown by dashed lines in left-hand panel) are in good agreement for low $\omega_0$ (corresponding to low $\sigma^2$), but start to diverge at high $\sigma^2$. Simulations use a physically viable log-normal over-density distribution (shown in right-hand panel), while the model assumes a Gaussian distribution (for which the pdf with matching $\sigma^2$ is shown as a solid curve in the right-hand panel). The further the underlying distribution strays from Gaussianity, the more inaccurate the model prediction. Realistic sky distributions will tend to be close to Gaussian.}
\label{fig:test_lognormal}
\end{figure*}

Another important assumption to check is that of Gaussianity of the underlying density field. It is well known that a Gaussian density field is physically inconsistent, since the support of the Gaussian distribution is over all real numbers, including negative numbers. A more physically appropriate distribution is the log-normal distribution \citep{Coles1991}. In over-density, $\delta_x$, this distribution has support on $[-1,\infty)$, and for a mean-zero field with small variance, is almost identical to a Gaussian distribution. Our simulations, produced with the \textsc{powerbox} package, use the log-normal distribution for this reason. However, producing a log-normal field requires a non-linear transformation which is not captured in our model prediction. Thus, in Fig. \ref{fig:test_lognormal} we show some comparisons between the inherently Gaussian model and the log-normal simulations.

We find, overall, that agreement is extremely good. The blue curves show consistency to an imperceptible level, as evidenced by the left-hand panel, while the underlying distribution clearly shows departure from Gaussianity. It is not until the high-variance green curve, with $u_0=0.9$, that inaccuracies become apparent, and the underlying field in this case is dramatically non-Gaussian. In practice, the sky should be reasonably Gaussian, as it is the integrated spectrum over many redshifts which are not fully correlated. Thus our Gaussian model is an excellent approximation.

\begin{figure*}
	\centering
	\includegraphics[width=0.9\linewidth]{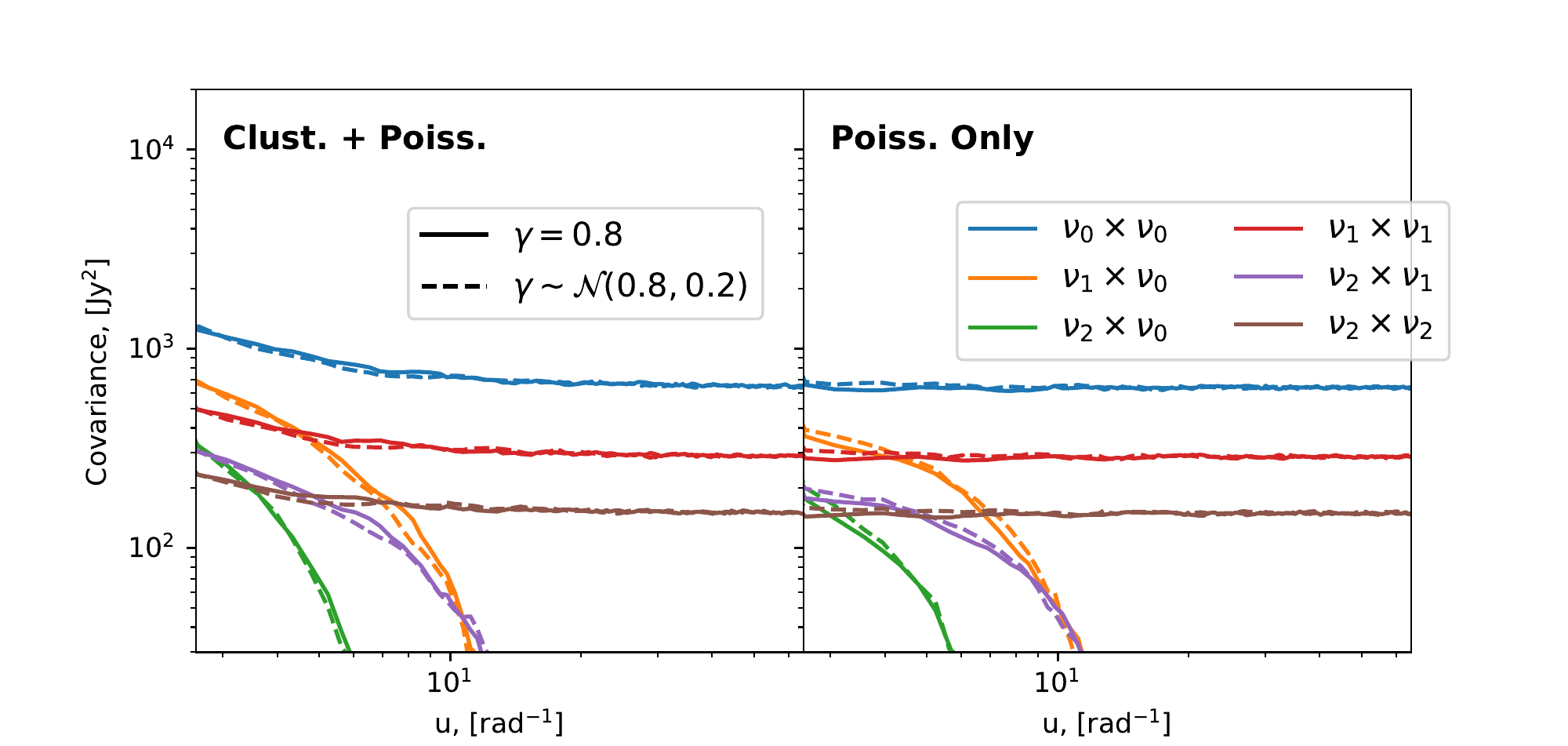}
	\caption{Comparison of simulated covariance with and without a universal SED. Universal SED parameters are identical to figure \ref{fig:compare_clustering_only}, whilst the non-universal SED comprises power-law SE's with slopes drawn from a normal distribution about $\gamma=0.8$ with width 0.2. No significant variation is evident.}
	\label{fig:spec_index_dist}
\end{figure*}

Our final test is to assess the impact of a non-universal SED.
Figure \ref{fig:spec_index_dist} shows the results of the covariance as a function of $u$, similarly to figure \ref{fig:compare_clustering_only}, but in which the solid lines use our fiducial universal SED, while the dashed lines use a non-universal slope for the underlying power-law SED. 
The distribution of $\gamma$ is taken from \cite{Callingham2017}, and is a relatively tight Gaussian distribution centred on $\gamma=0.8$, with a width of 0.2. 
The non-universal SED simulation was run such that the total number of sources at each frequency was constant, which mimics actual observations.
The figure shows clearly that the disparity is minimal, certainly negligible compared to the uncertainty in the model parameters.
While it may be interesting to determine the limits on variability of the SED under a number of scenarios, we do not further pursue it in this work, since clearly realistic models of the variability are inconsequential.

\subsubsection{Model Features in Fourier Space}
\label{sec:model_fourier}

The quantity that is directly used in the CHIPS algorithm is not the covariance as a function of frequency, $\vect{C}_{\rm FG}$, but its transformation to the covariance of the 2D power modes $\vect{C}_P$ (cf. Eq. \ref{eq:power_covariance}). 
We note that though in principle the CHIPS formalism utilises the full covariance matrix in Fourier space, the remaining plots will show only the variance, $\sigma^2_P(\kpar,k_\perp) = \delta(\kpar'-\kpar'') \vect{C}(\kpar',\kpar'',k_\perp)$.

\begin{figure*}
\centering
\includegraphics[width=0.9\linewidth]{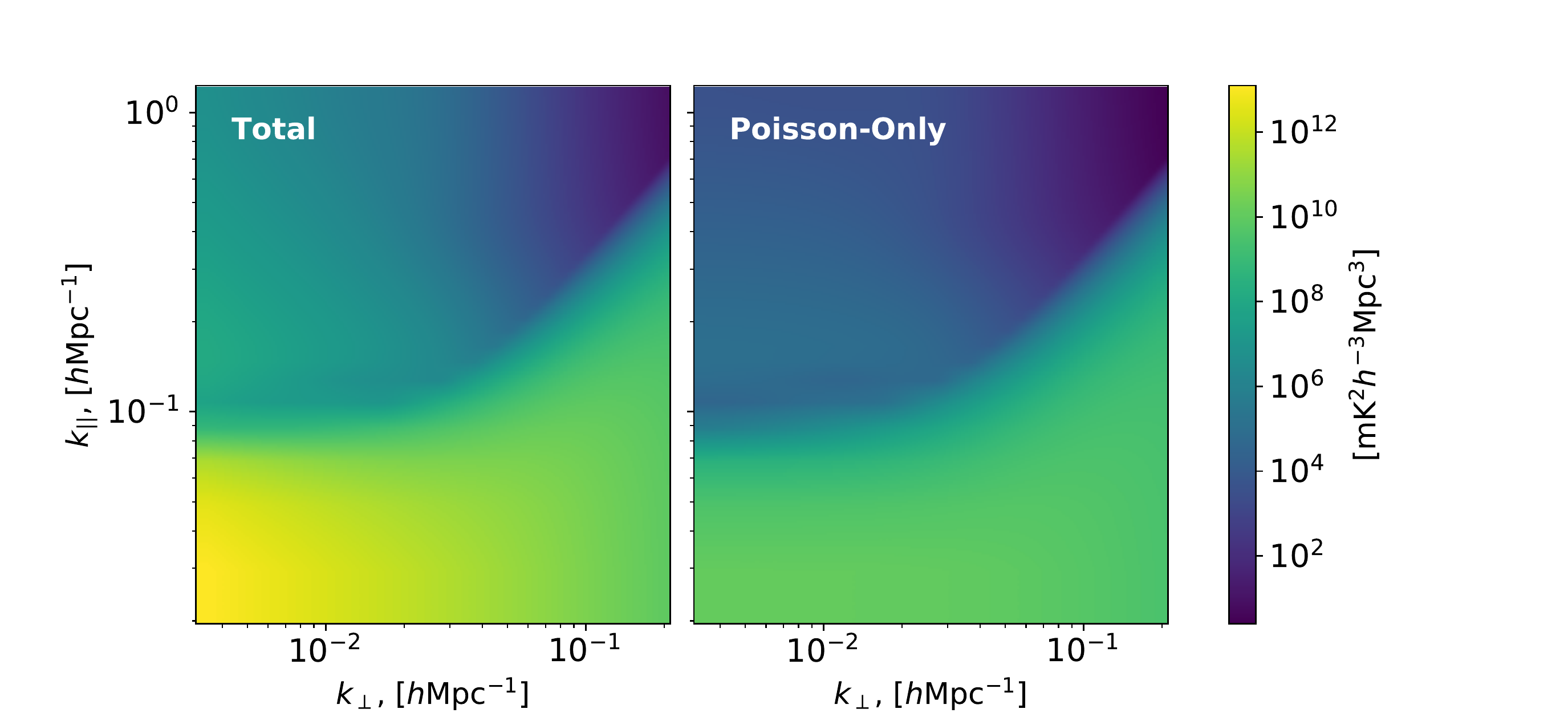}
\caption{Standard deviation of the power in 2D modes, for both the total clustering Poisson solution (left-hand panel) and Poisson-only solution (right-hand panel) using a fiducial (MWA-like) model (see Table \ref{tab:parameters}). Colour-scales are equivalent in each panel to aid comparison.}
\label{fig:compare_total_poisson_PS}
\end{figure*}

In Fig. \ref{fig:compare_total_poisson_PS}, we show $\sigma_P$ for each 2D mode of the power spectrum, for both the clustering and Poisson-only solution (left and right panels respectively). This quantity is closely related to the foreground-contributed power, hence the likeness to the standard 2D PS plots, with clearly visible wedge feature. There is a noticeable excess of power at low $k_\perp$ in the total solution with clustering, both in the foreground-dominated region and the window. This arises from the excess large-scale power contributed by source clustering.

A measurement of interest is the relative importance of the clustering term at any 2D mode. 
After all, if the clustering term is typically sub-dominant, it may not be worth the additional computational and intellectual resources required to include it in our models. 

%A more illustrative way to see the effects of the improved solution is to view the ratio of the clustering term to the total foreground PS, $P_{\rm Clust}/(P_{\rm Clust} + P_{\rm Poiss})$. This quantity can be interpreted in two ways: either as the fractional contribution of the clustering term (i.e. its importance), or as the bias introduced by ignoring the clustering relative to the overall foreground power.

\begin{figure*}
\centering
\includegraphics[width=0.9\linewidth,trim=0cm 0cm 2cm 0cm]{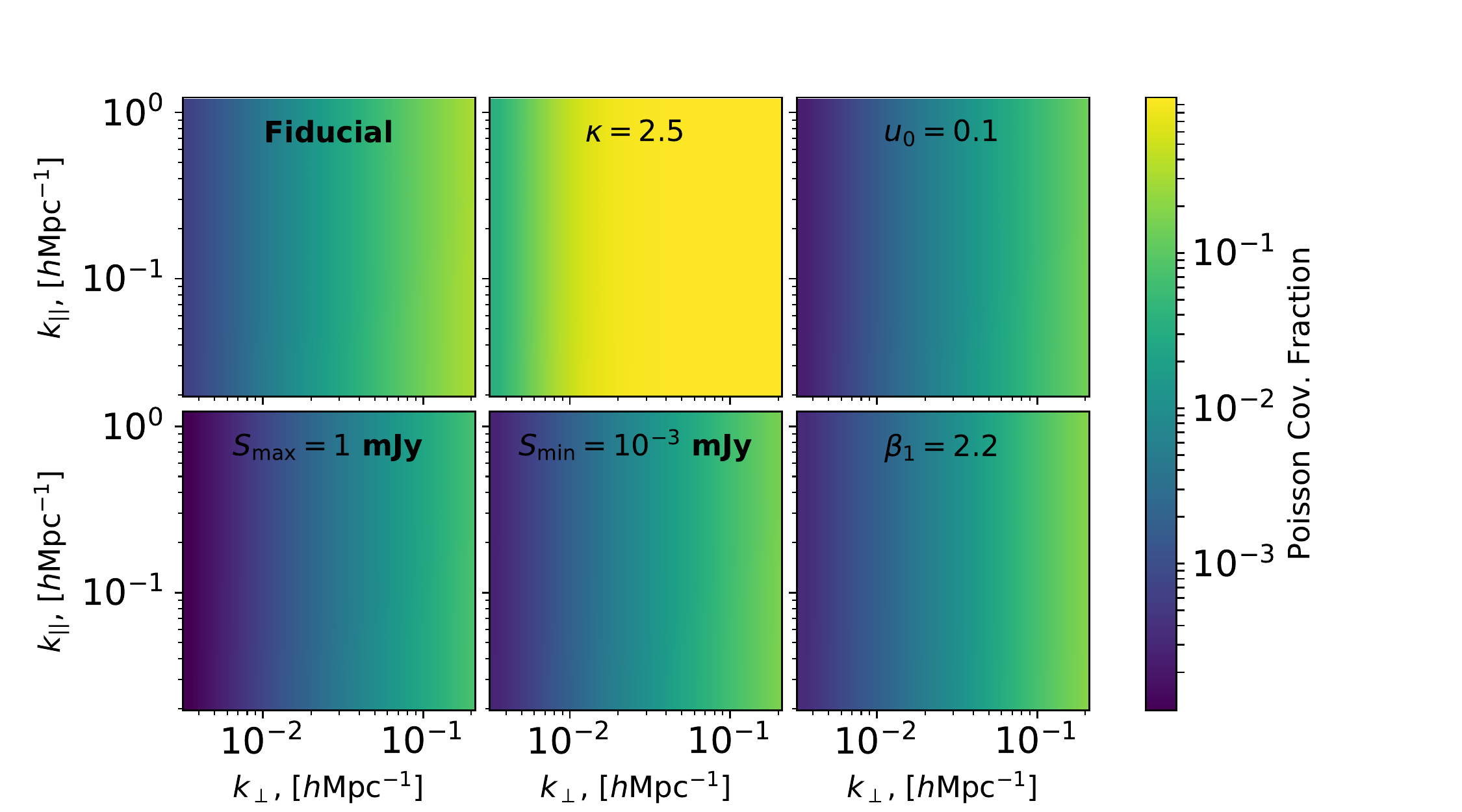}
\caption{Survey of the effects of various parameters of the model on the ratio of the Poisson-only to the total 2D PS. The top-left panel is the fiducial case, and moving clockwise, the slope, the peeling flux density limit, and the power spectrum normalisation are modified.}
\label{fig:compare_ratios_bias_PS}
\end{figure*}

In Fig. \ref{fig:compare_ratios_bias_PS} we display the quantity $\sigma_{\rm Poiss}/(\sigma_{\rm Clust} + \sigma_{\rm Poiss})$ -- i.e. the relative contribution of the Poisson standard deviation to the total, for several sets of parameters. The fiducial case is the same as used above. 

The first point to note is that the clustering fraction is almost exclusively dependent on $k_\perp$, with negligible dependence on $\kpar$.
The form of this dependence is such that the clustering is dominant at small $k_\perp$, and diminishes towards smaller scales. The point at which it reaches a fraction of 0.5 will be closely related to $u_\star$.

We are here concerned primarily with the effects of the parameters, not the actual ratio for any given set. Thus we see that an increase in the power-spectral slope results in a compression of foreground power into very large-scale modes, and potentially out of the observation altogether. Alternatively, an increase in $u_0$, or reduction of $S_{\rm min}$ or $S_{\rm max}$ has the opposite effect of increasing the range of clustering-dominance. The slope of the faint source counts does not seem to have a large impact for the default parameters employed here.

%Thus the two panels showing the effect of modifying the slope and normalisation of the input point-source PS are easily interpreted -- the normalisation merely affects the overall normalisation of the 2D PS, while the slope accordingly affects the slope of the $k_\perp$ dependence. 
%Of some interest is the effect of modifying the peeling limit. For the simple power-law source count model employed here, a reduction of the peeling limit (which corresponds to the expected situation with new telescopes coming online) \textit{enhances} the effect of clustering, resulting in a more pronounced ratio at small $k_\perp$. 

\section{Application of Clustering Model to Synthetic EoR Data}
\label{sec:application}
To quantify the importance of including clustering information in the visibility covariance, we perform a mock test, by measuring the PS from synthetic data, with clustered foregrounds included, using both the original and improved solutions.

\subsection{Simulation Details}
We use the \textsc{faint galaxies} simulation from the \textit{Evolution of Structure} (EOS) project\footnote{\url{http://homepage.sns.it/mesinger/EOS.html}}, which was run with \textsc{21cmFAST} \citep{Mesinger2011}. 
The simulation has a size of 1600 cMpc, with 1024$^3$ grid cells, and is provided as a `best guess' at reionization physics. 
We specifically use the brightness temperature lightcone which covers our redshift range of interest. 
The $z$-axis co-ordinates are converted to redshifts by assuming constant intervals in comoving distance under the default Planck15 cosmology, and the box is cut to contain the range of frequencies 150 - 161 MHz in 128 bins, approximating an MWA observation.

Ideally, one would process the simulation through an instrumental pipeline, so that all instrumental biases present in the covariances would also be present in the synthetic signal. 
However, the angular size of the simulation is too small at a redshift of $\sim 9$ to cover a single primary beam width, which undermines this approach. 
Instead, we choose to simply measure the 2D signal by 3D Fourier-transforming the original box and averaging over the perpendicular modes of the (volume-normalized) power spectrum.
%Figure \ref{fig:signaltonoiselowkappa} shows the 2D signal arising from this process.
%As expected, the signal takes the form of a smooth close-to-isotropic power-law-like function diminishing towards small scales.
Because of this the perpendicular scales covered span a large portion of those available to the MWA (and planned for SKA1-LOW), but do not reach to the largest scales we can probe.
This should be kept in mind for the remainder of this section, as the largest influence of the clustering solution arises on these large scales.

\subsection{Signal-to-noise estimates}
\begin{figure*}
    \centering
    \includegraphics[width=0.85\linewidth, trim=1cm 1cm 2cm 2cm]{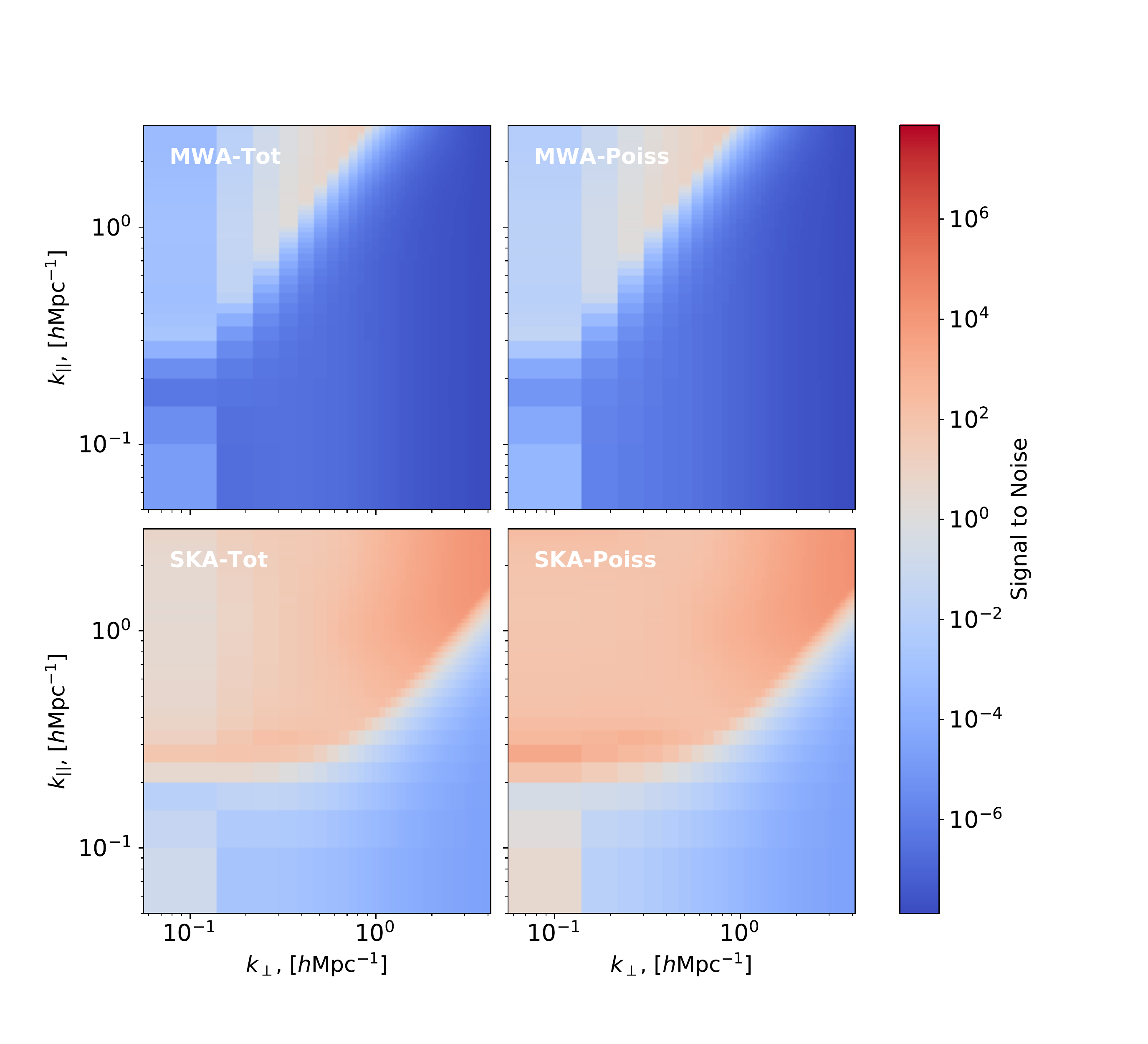}
    \caption{Signal-to-noise ratios for a signal derived from the simulation described in text, and foreground models based on table \ref{tab:parameters}. Top panel and bottom panels use MWA-like and SKA-like parameters respectively, while left- and right-hand panels indicate full and Poisson-only solutions.}
    \label{fig:signaltonoiselowkappa}
\end{figure*}

Figure \ref{fig:signaltonoiselowkappa} shows the signal-to-noise, defined as $P_{\rm signal}/\sigma(k_\perp, \kpar)$, for a variety of realisations of the foreground model.
The basic parameters of the model retain their fiducial values (cf. Table \ref{tab:parameters}), while the top panel uses the MWA-like parameters where appropriate, and the lower panel uses SKA-like parameters. 
Left- and right-hand panels again indicate full and Poisson-only solutions. 

We remind the reader that only extra-galactic point-source foregrounds are included in this plot. 
Thus it represents what would be achieved with an infinite amount of observing time with a well-known instrument, after perfect removal of Galactic foregrounds and other systematics. 
Clearly, this should not be taken as a prediction for a detection or otherwise using our model, but rather we are interested in the \textit{relative} behavior of our model compared to the Poisson solution, with respect to a fiducial signal. 

The most striking feature is that the SKA performs significantly better than the MWA. 
This is not directly attributable to the instrument, but rather to the level of peeling performed in the experiment. 
Nevertheless, this clearly relates back to the instrument's capability to deeply model individual foreground sources to provide a catalogue for this peeling.

For both MWA- and SKA-like mock observations, the lowest $k_\perp$ modes have a significantly reduced signal-to-noise in the full solution as compared to the Poisson-only solution. 
This is just as expected, due to the extra noise from the source PS.
We do note however that these high $k_\perp$ window-modes already have a suppression of signal-to-noise even in the Poisson case.
This will limit the influence of the clustering term in the final averaged 1D PS. 

\subsection{Effects in the 1D PS}
The ultimate objective is to determine the effects of ignoring the clustering term on the 1D PS. 
We expect two effects may arise -- a bias from ignoring extra foreground power, and an artificially decreased uncertainty. 

The CHIPS formalism does not remove the foreground power from the estimated signal due to the potential for over-subtraction of the dominant source of power. 
Instead, since the goal at this point is merely for a robust \textit{detection}, rather than an accurate \textit{measurement} of the signal, it leaves the foregrounds in the estimation, checking whether they are consistent with non-zero power through the ascribed uncertainty. 
Thus, we generate our mock estimated 1D PS via
\begin{equation}
    P(k_i) = \frac{1}{\sum_j 1/\sigma^2_{,i}} \sum_j \frac{P_{,i} + \sigma_{,i}}{\sigma^2_{,i}},
\end{equation}  
where the ${}_{,i}$ notation indicates that we are using entries only in the $i^{th}$ radial bin. 
Furthermore, the uncertainty is estimated as the weighted average across the bin:
\begin{equation}
    \sigma^2(k_i) = \frac{N_i}{\sum_j 1/\sigma^2_{,i}}.
\end{equation}

\begin{figure*}
    \centering
    \includegraphics[width=\linewidth, trim=1cm 0cm 1cm 0cm]{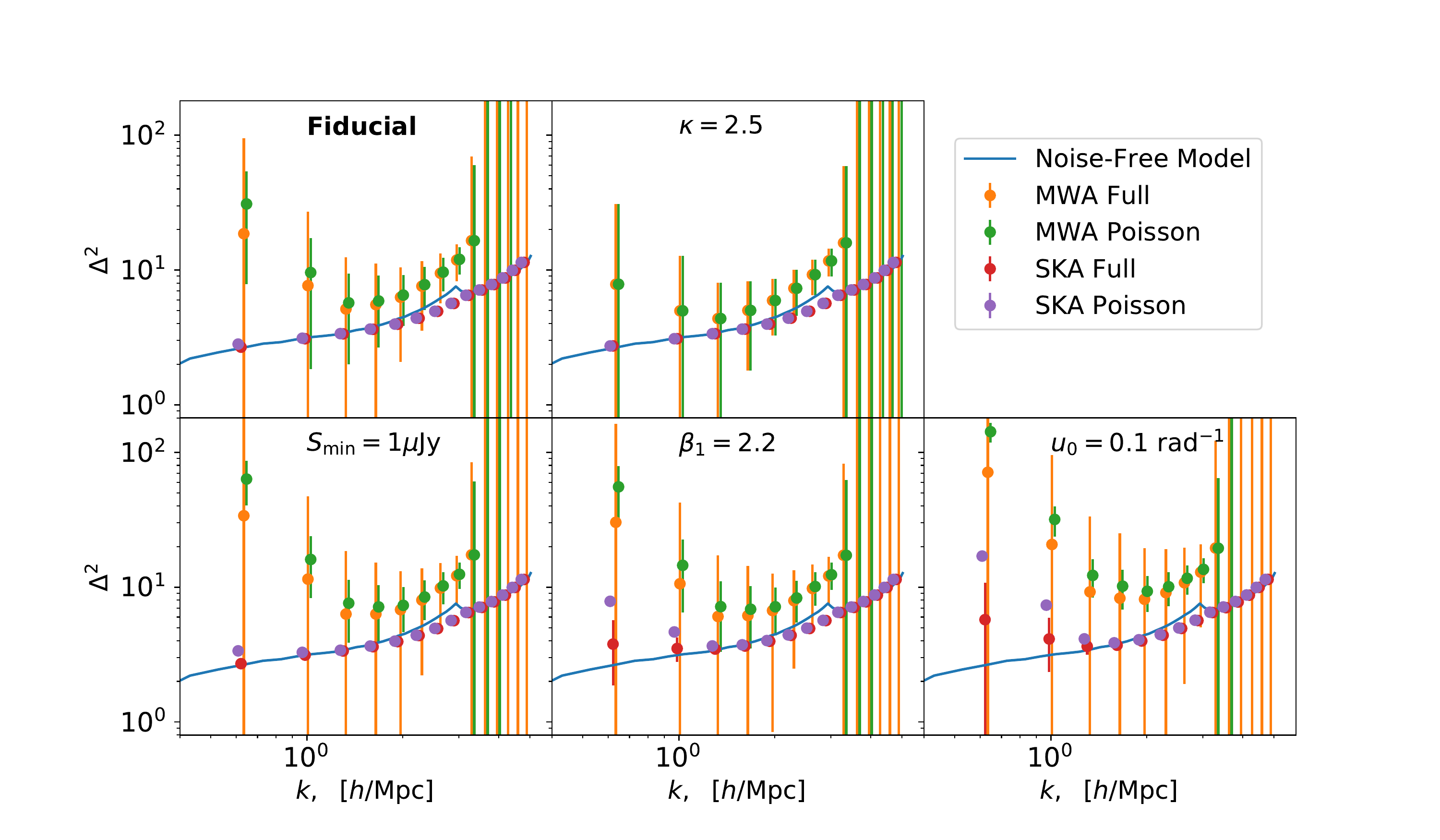}
    \caption{Inverse-covariance weighted 1D power spectra. Each panel shows MWA-like and SKA-like mock observations, for both the full solution and the Poisson-only solution, as well as the underlying signal. Different panels involve different model parameter choices. Error-bars are derived in text.}
    \label{fig:final1dplotspanels}
\end{figure*}

Figure \ref{fig:final1dplotspanels} shows the results of this averaging over the 2D spectra shown in Fig. \ref{fig:signaltonoiselowkappa} (top-left panel) and several other parameter choices.
Again, results here are not to be interpreted as actual predictions, because of the lack of other foregrounds, but we do gain insight as to the importance of including the clustering information.

For most parameter choices, the SKA results in a very precise and accurate measurement of the input signal -- as compared to the MWA model which always includes a significant bias.
Again, this is due to the level of peeling performed in this hypothetical experiment.
Nevertheless a few interesting features arise in the MWA-like models.
Firstly, in most cases presented here, there is discernible extra bias if the clustering term is ignored.
In general, the amplitude of this extra bias is increased for models in which faint sources are relatively more abundant, or where the source spectrum is of higher amplitude.  
The opposite is true when the source power spectrum is very steep. In this case, the clustering plays no visible role.

At the same time, as expected, the error-bars are artificially reduced when ignoring clustering. 
Importantly, in almost all cases (except the steep-source-spectrum case) there are scales at which a \textit{false detection} occurs if clustering is ignored. 
That is, there are scales at which the bias increase and artificial decrease in uncertainty would indicate a significant non-zero power, when in fact the full model shows that no detection should be attainable.
Indeed, if faint sources are highly abundant and the source spectrum is shallow (lower centre and right-hand plots), even the fiducial SKA model yields significantly inaccurate measurements\footnote{We note that the mid-$k$ bias in all MWA models is primarily a result of anisotropy of the signal due to redshift space effects, rather than a fault in the model itself \citep[cf.][]{Jensen2016}.}.

This is strong motivation for including our clustering extension in the general CHIPS framework in the future. 
Besides this, there is strong motivation for future surveys to constrain the model parameters -- especially the faint source count slope $\beta_1$, and the source power spectrum parameters $u_0$ and $\kappa$. 
We note that surveys at higher frequencies have measured these quantities \citep[eg.]{Blake2004}, but how they translate to lower frequencies and fainter sources is uncertain.
There is already work underway to measure these quantities with low-frequency surveys such as GLEAM and EoR-specific deep pointings \citep{Hurley-Walker2016}.

\section{Conclusions and Future Prospects}
\label{sec:conclusion}
In this paper we have derived an improved statistical point-source foreground covariance model for use in inverse-covariance suppression schemes in the context of the EoR. 
In particular, beginning from the basic CHIPS model proposed in \cite{Trott2016}, we investigated extensions to two components. We extended: (i) the source count distribution from a single power-law to an arbitrarily broken power-law, and (ii) the spatial distribution from a uniform Poisson-process to an arbitrary isotropic distribution -- especially a power-law configuration.  

In doing so, we have derived a general framework for including these extensions, applicable in principle to current and future interferometers. 
In addition we have verified the validity of our expressions by comparison to statistical simulations, along with providing an expression for the on-sky scale above which our extensions begin to dominate the covariance.

\subsection{Future Prospects}
There are a number of ways in which the present analysis may be improved and extended.
In terms of improvement, we have noted that the framework sits more naturally within the context of spherical harmonics, which would dispense with several of the approximations made in this paper \citep[eg.][]{Liu2016}. 
Furthermore, we have maintained the simplistic assumption of complete uniformity of $S(\nu)$ across sources -- an assumption that is surely inaccurate. 
It remains to be seen how strong an effect a relaxation of this assumption would incur, though we noted that it could alter the result that all source-count model effects are sequestered into a single factor. 
A more drastic improvement would be to include covariance between baselines in the framework.
At this point, we have considered only covariance between frequencies for the same baseline, which greatly simplifies the calculations.

As far as extensions of this framework are concerned, a useful idea may be to link it with a physically-motivated prescription for the source population models.
An example of this may be a radio-frequency conditional luminosity function \cite[CLF;][]{Cooray2006} as a function of redshift, out of which the source counts, SED and source clustering may be derived. 
This would simultaneously provide a convenient way to model the foregrounds for suppression, but also to understand them in their own right.

We also note that the general approach to calculate the clustering term for the covariance presented here may be utilised in other foreground regimes.
In particular, diffuse Galactic emission is often described as isotropic turbulence within a single observed field, which may be succinctly quantified under the same framework. 
Furthermore, the ionosphere's \textit{total electron content} (TEC) is typically prescribed as near-Kolmogorov turbulence (with the potential addition of a large-amplitude single wave-mode to describe travelling ionospheric disturbances).
In this case, the TEC field acts as a differential screen, so that the resulting foreground covariance must be derived via a similar approach as that used in weak lensing studies.
Nevertheless, it is easy to see that even this contamination source can be connected to the formalism presented here.

\subsection{Conclusions}
Predictions of our model extension can be summarised as
\begin{enumerate}
    \item There is excess foreground power at large $k_\perp$;
    \item This excess foreground is a function almost exclusively of $k_\perp$, rather than $\kpar$;
    \item Models with a higher relative abundance of faint sources in general \textit{increase} the relative impact of ignoring our extensions;
    \item Models with steeper source spectra \textit{decrease} the relative impact of ignoring our extensions;
    \item Under some parameter choices, ignoring the extensions induces \textit{false detections}, i.e. a significant non-zero power which in fact is merely bias.
\end{enumerate}

These predictions indicate that future measurements of the EoR PS could be greatly benefited from including these model extensions -- particularly the clustering of sources. 
Furthermore, while deeper surveys will render the total point-source foreground contamination less important, the remainder of that contamination will be more dominated by clustering. 
Thus using SKA1-LOW in generating accurate \textit{measurements} of the EoR PS will require taking into account the framework we have presented here.

These results also motivate the accurate determination of source count models down to faint limits, as well as point-source power spectra. 
Future deep surveys such as GLEAM-X \citep{Hurley-Walker2017} and those performed with SKA1-LOW will be indispensable in providing high fidelity measurements of these models, which will feed back into the covariance models for the EoR.

\section*{Acknowledgments}
Parts of this research were conducted by the Australian Research Council Centre of Excellence for All-sky Astrophysics (CAASTRO), through project number CE110001020.
SGM would like to thank Jean-Pierre Macquart, Randall Wayth and Cullan Howlett for useful discussions throughout the preparation of this paper. 
This research has made use of NASA's Astrophysics Data System.
CMT is supported under the Australian Research Council's Discovery Early Career Researcher funding scheme (project number DE140100316). This work was supported by resources provided by the Pawsey Supercomputing Centre with funding from the Australian Government and the Government of Western Australia. We acknowledge the International Centre for Radio Astronomy Research (ICRAR), a Joint Venture of Curtin University and The University of Western Australia, funded by the Western Australian State government.

\bibliography{Mendeley}
\begin{appendix}

\section{Cosmological Unit Conversions}
\label{app:cosmo}
Conversion between observation parameters, $(\vect{u},\nu)$ and the cosmological parameters is given by \cite{Morales2004} as
\begin{align}
k_\perp &= \left|\frac{2\pi \vect{u}}{D_M(z)} \right|   \label{eq:kperp_to_u}\\
\kpar &= \frac{1}{\nu} \frac{H_0 f_{21}E(z)}{c(1+z)^2} \equiv \frac{G(z)}{\nu}. \label{eq:hz_to_mpc}
\end{align}
Here $z$ is the central redshift of the observation, $f_{21}\sim 1.42$GHz is the rest frequency of the 21 cm line, $D_M(z)$ is the transverse comoving distance at that redshift, and in a flat universe such as we adopt
\begin{equation}
E(z) = \sqrt{\Omega_m(1+z)^3 + \Omega_\Lambda}.
\end{equation} 

The natural units of the power spectrum are ${\rm Jy^2}\,{\rm Hz}^2$, but these can be converted to the standard units of ${\rm mK}\,{\rm Mpc}^3\,h^{-3}$ using the following recipe. To convert Hz to Mpc$/h$, one divides by $G(z)$, then to convert Jy to K.sr, we use the following conversion:
\begin{equation}
{\rm K.sr} = {\rm Jy}\frac{A_{\rm eff}}{2\times 10^{26}k_B}, 
\end{equation}
where $A_{\rm eff}\sim 20 {\rm m}^2$ is the effective area of the MWA telescope at $\nu\sim 150$ MHz, and $k_B$ is Boltzmann's constant. The steradians are converted to Mpc$^2\,h^{-2}$ via a squared factor of the comoving distance, $D^2(z)$, and finally the value is normalised by the volume:
\begin{equation}
V = \frac{c^2}{A_{\rm eff}\nu_{\rm max}}\Delta\nu \frac{D^2(z)}{G(z)},
\end{equation} 
where $\Delta\nu$ is the range of frequencies in the observation. Explicitly, the conversion from ${\rm Jy^2}{\rm Hz}^2$ to ${\rm mK}\,{\rm Mpc}^3\,h^{-3}$ is
\begin{equation}
{\rm mK}\,{\rm Mpc}^3\,h^{-3} = {\rm Jy^2}\,{\rm Hz}^2 \frac{A_{\rm eff}D^2(z) \nu^2_{\rm max}}{c^2\Delta\nu G(z) 2k_B 10^{26}}.
\end{equation}

\end{appendix}

\end{document}